\documentclass[preprint,showpacs,preprintnumbers,amsmath,amssymb]{revtex4}

\pdfoutput=1

\usepackage{amsmath}  
\usepackage{amsfonts} 
\usepackage{graphicx} 

\begin{document}

\title{Bound states of the one-dimensional Dirac equation \\
for scalar and vector double square-well potentials
}

\date{\today}

\begin{abstract}
We have analytically studied bound states of the one-dimensional Dirac equation  
for scalar and vector double square-well potentials (DSPs),
by using the transfer-matrix method.
Detailed numerical calculations of the eigenvalue, wave function and density probability
have been performed for the three cases: (1) vector DSP only, (2) scalar DSP only, 
and (3) scalar and vector DSPs with equal magnitudes.
We discuss the difference and similarity among results of the cases (1)-(3)
in the Dirac equation and that in the Schr\"{o}dinger equation.
Motion of a wave packet is calculated for a study on
quantum tunneling through the central barrier in the DSP.
\end{abstract}


\pacs{03.65.Ge, 03.65.Pm, 31.30.jx}

\maketitle 

\section{Introduction}
The basic physics of relativistic quantum mechanics was formulated in the Dirac equation, 
which elucidates the origin of spin 1/2 of an electron
and predicts the existence of an antiparticle (a positron) \cite{Greiner81}.
The Dirac equation has been applied not only to realistic models like hydrogen atom
but also to pedagogical models which play important roles in understanding 
the properties of the Dirac equation. 
The Dirac equation for step and square potentials has been investigated 
in connection to the Klein paradox \cite{Thomson91,Nitta99,Krekora04,Le06,Bosanac07}. 
Square-well potentials with finite and infinite depths 
have been studied in Refs. \cite{Coulter71,Gumbs86,Coutinho88,Alberto96,Alhaidari11,Ying04}.
The double square-well potential (DSP) consisting of the confining potential
and the central potential is more difficult than the single square-well potential 
\cite{Coulter71,Gumbs86,Coutinho88,Alberto96,Alhaidari11,Ying04}.
Indeed applications of the Dirac equation to the DSP 
have not been reported as far as we are aware of \cite{Note1}.
The DSP is a simplified model for an appropriate and realistic description
of a continuous double-well potential.
Extensive investigations within the nonrelativistic treatment of the Schr\"{o}dinger equation
have been made for double-well systems where numerous quantum phenomena have been realized
(for a recent review on double-well systems, see Ref. \cite{Thorwart01}). 
The Schr\"{o}dinger equation for the DSP with the infinite confining potential 
is manageable and treated in the undergraduate text, 
whereas the DSP with the {\it finite} confining potential has been investigated 
only in several studies \cite{Lopez06,Acus11,Tserkis13}.
One of advantages of the DSP is to provide us with exact analytic expressions 
for eigenstates and wave functions.
In the relativistic quantum theory, two types of vector ($V(x)$) 
and scalar ($S(x)$) potentials have been adopted.
In previous studies on the single square-well potential, the vector potential was adopted
in Refs. \cite{Coulter71,Gumbs86,Coutinho88,Alhaidari11,Ying04}
while the scalar potential was employed in Refs. \cite{Coutinho88,Alberto96}.
The purpose of this paper is to make a detailed study on 
the Dirac equation for scalar and vector DSPs and to make a comparison 
between results of the Dirac equation and the Schr\"{o}dinger equation.
Such a study is expected to be essential and inevitable for a deeper understanding of 
relativistic quantum double-well systems.

The paper is organized as follows.
In Sec. II, we obtain analytic expressions for eigenvalues 
and wave functions of bound states in the Dirac equation for scalar and vector DSPs,
by using the transfer-matrix method. In Sec. III, the transcendental complex equation
for the eigenvalue is numerically solved and bound-state wave functions are obtained
for three cases: (1) the vector DSP only (VDSP: $S(x)=0$),
(2) the scalar DSP only (SDSP: $V(x)=0$), and
(3) equal scalar and vector DSPs (EDSP: $S(x)=V(x)$).
In Sec. IV, eigenvalues in the Dirac equation are compared
to those obtained in the Schr\"{o}dinger equation.
Motion of a wave packet is investigated for a study on the quantum tunneling 
through the central barrier in the DSP.  
Sec. V is devoted to our conclusion.
In the Appendix the transfer-matrix method 
is applied to the Schr\"{o}dinger equation for the DSP.

\section{Dirac equation for the double square-well potential}
\subsection{Transfer-matrix formulation}
We will obtain the bound-state solution of the stationary one-dimensional Dirac equation 
for the DSP.
Among conceivable, equivalent expressions for the Dirac equation, 
we employ the $(1+1)$-dimensional representation 
\begin{eqnarray}
\left[c \:\sigma_x \left( -i \hbar \frac{\partial}{\partial x} \right)
+ \sigma_z \left[m c^2+S(x) \right] \right] \Psi(x) &=& \left[E-V(x) \right] \:\Psi(x), 
\label{eq:A2}
\end{eqnarray}
with
\begin{eqnarray}
\Psi(x) &=& \left( {\begin{array}{*{20}c}
   \psi_+(x)  \\
   \psi_-(x)   \\
\end{array}} \right),
\label{eq:A3}
\end{eqnarray}
where $\psi_{\pm}(x)$ signify elements of two-dimensional spinor of $\Psi(x)$,
$\sigma_x$ and $\sigma_z$ are Pauli matrices,
$S(x)$ and $V(x)$ express scalar and vector potentials, respectively,
$E$ denotes the stationary energy, $m$ is rest mass of a particle with spin 1/2, 
and $c$ is the light velocity.
Two component of $\psi_+(x)$ and $\psi_-(x)$ satisfy
\begin{eqnarray}
[mc^2+V(x)+S(x)] \psi_+(x)- i \hbar c \frac{d}{dx} \psi_-(x)&=& E \psi_+(x), 
\label{eq:A4} \\
- i \hbar c \frac{d}{dx} \psi_+(x)+[-mc^2+V(x)-S(x)] \psi_-(x)&=& E \psi_-(x).
\label{eq:A5}
\end{eqnarray}

\begin{figure}
\begin{center}
\includegraphics[keepaspectratio=true,width=60mm]{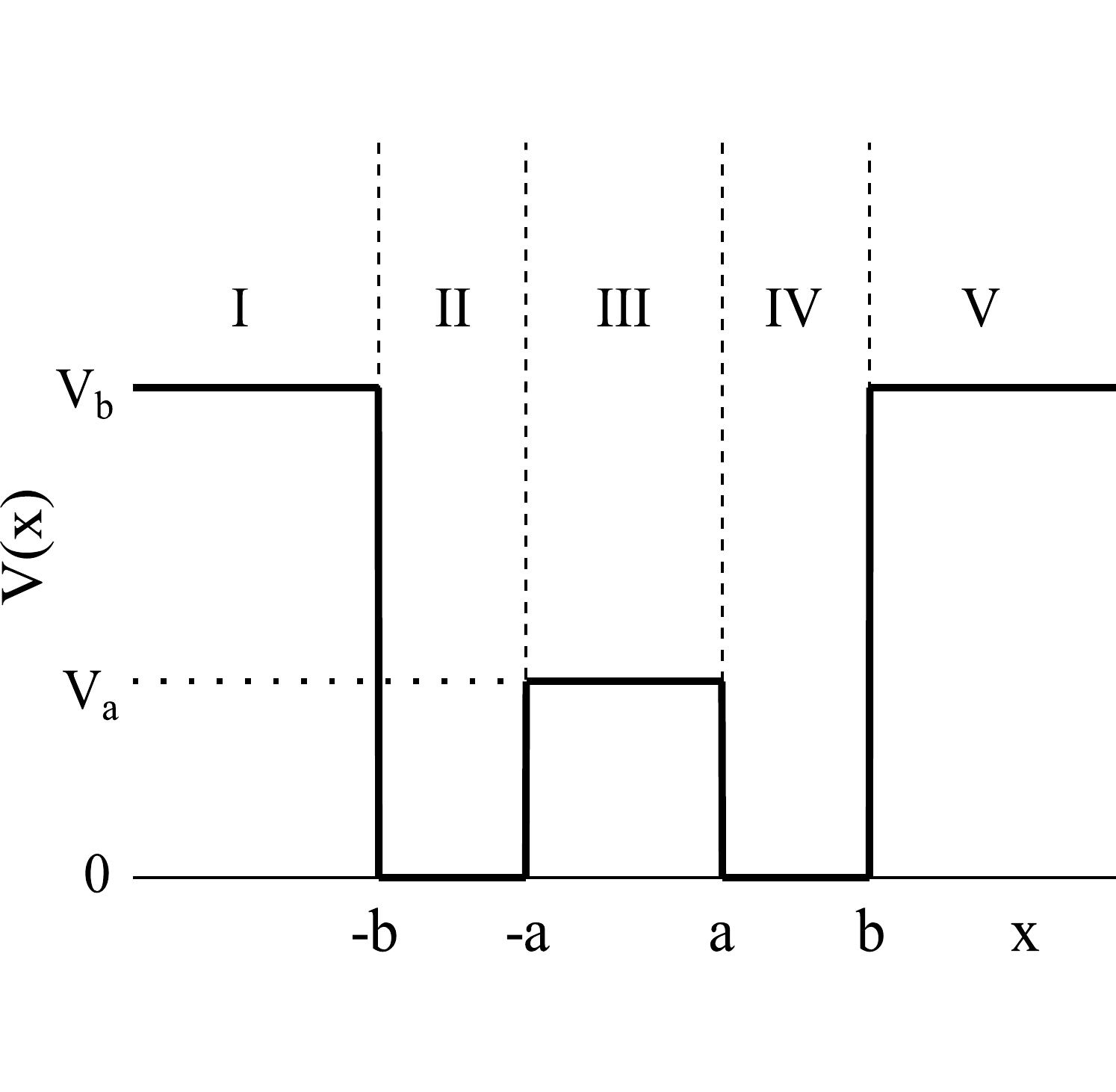}
\end{center}
\caption{
Schematic vector DSP, $V(x)$, given by Eqs. (\ref{eq:A1}) (bold solid lines),
the $x$ axis being divided into five regions I-V separated by dashed lines.
The scalar DSP, $S(x)$, is given if we read $V_a \rightarrow S_a$ and 
$V_b \rightarrow S_b$.
}
\label{fig1}
\end{figure}

We consider the one-dimensional vector potential $V(x)$ expressed by
\begin{eqnarray}
V(x) &=& \left\{ \begin{array}{ll}
V_b
\quad & \mbox{for $x \leq -b $ \hspace{1.2cm} (region I)}, \\ 
0
\quad & \mbox{for $-b < x \leq -a $ \hspace{0.2cm} (region II)}, \\
V_a
\quad & \mbox{for $-a < x \leq a $ \hspace{0.5cm} (region III)}, \\ 
0
\quad & \mbox{for $a < x \leq b $ \hspace{1cm} (region IV)}, \\
V_b
\quad & \mbox{for $x > b $ \hspace{1.5cm} (region V)},
\end{array} \right. 
\label{eq:A1}
\end{eqnarray}
with $V_b \geq 0$ and $0 \leq V_a \leq V_b$.
Here the $x$ axis is divided into five spatial regions: 
(I) $x \leq -b$, (II) $-b \leq x \leq -a$, (III) $-a < x \leq a$, 
(IV) $a< x \leq b$, and (V) $x > b$;
$V_b$ expresses the confining potential in the regions I and V;
$V_a$ denotes central barrier potential in the region III (Fig. \ref{fig1}).

As for the scalar potential $S(x)$, we consider
\begin{eqnarray}
S(x) &=& \left\{ \begin{array}{ll}
S_b
\quad & \mbox{for $x \leq -b $ \hspace{1.2cm} (region I)}, \\ 
0
\quad & \mbox{for $-b < x \leq -a $ \hspace{0.2cm} (region II)}, \\
S_a
\quad & \mbox{for $-a < x \leq a $ \hspace{0.5cm} (region III)}, \\ 
0
\quad & \mbox{for $a < x \leq b $ \hspace{1cm} (region IV)}, \\
S_b
\quad & \mbox{for $x > b $ \hspace{1.5cm} (region V)},
\end{array} \right. 
\label{eq:A1b}
\end{eqnarray}
with $S_b \geq 0$ and $0 \leq S_a \leq S_b$ (read $V_a \rightarrow S_a$
and $V_b \rightarrow S_b$ in Fig. \ref{fig1}).
The adopted scalar and vector DSPs are symmetric with respect to the origin. 
In the limit of $V_a=S_a=0$, $a=0$, or $a=b$, the double square-well potential 
reduces to the single one.

Wave functions in five regions I-V may be expressed by
\begin{eqnarray}
\Psi_I(x)  &=& A_1 \left( {\begin{array}{*{20}c}
   1  \\
   \beta   \\
\end{array}} \right)e^{iqx}  + B_1 \left( {\begin{array}{*{20}c}
   1  \\
   { - \beta }  \\
\end{array}} \right)e^{ - iqx} 
\hspace{1cm}\mbox{for $x < -b$},
\label{eq:A6}
\end{eqnarray}
\begin{eqnarray}
\Psi _{II}(x)  &=& A_2 \left( {\begin{array}{*{20}c}
   1  \\
   \alpha   \\
\end{array}} \right)e^{ikx}  + B_2 \left( {\begin{array}{*{20}c}
   1  \\
   { - \alpha }  \\
\end{array}} \right)e^{ - ikx} 
\hspace{1cm}\mbox{for $-b < x < - a$},
\label{eq:A7}
\end{eqnarray}
\begin{eqnarray}
\Psi _{III}(x)  &=& A_3 \left( {\begin{array}{*{20}c}
   1  \\
   \gamma   \\
\end{array}} \right)e^{ipx}  + B_3 \left( {\begin{array}{*{20}c}
   1  \\
   { - \gamma }  \\
\end{array}} \right)e^{ - ipx} 
\hspace{1cm}\mbox{for $-a < x < a$},
\label{eq:A8}
\end{eqnarray}
\begin{eqnarray}
\Psi _{IV}(x)  &=& A_4 \left( {\begin{array}{*{20}c}
   1  \\
   \alpha   \\
\end{array}} \right)e^{ikx}  + B_4 \left( {\begin{array}{*{20}c}
   1  \\
   { - \alpha }  \\
\end{array}} \right)e^{ - ikx} 
\hspace{1cm}\mbox{for $a < x < b$},
\label{eq:A9}
\end{eqnarray}

\begin{eqnarray}
\Psi _{V}(x)  &=& A_5 \left( {\begin{array}{*{20}c}
   1  \\
   \beta   \\
\end{array}} \right)e^{iqx}  + B_5 \left( {\begin{array}{*{20}c}
   1  \\
   { - \beta }  \\
\end{array}} \right)e^{ - iqx} 
\hspace{1cm}\mbox{for $x < -a$},
\label{eq:A10}
\end{eqnarray}
with
\begin{eqnarray}
k &=& \frac{\sqrt{E^2-m^2 c^4}}{\hbar c}, 
\label{eq:A11} \\
p &=& \frac{\sqrt{(E+mc^2-V_a+S_a)(E-m c^2-V_a-S_a)}}{\hbar c}, 
\label{eq:A12} \\
q &=& \frac{\sqrt{(E+mc^2-V_b+S_b)(E-m c^2-V_b-S_b)}}{\hbar c}, 
\label{eq:A13} \\
\alpha &=& \frac{\hbar c k}{E+m c^2}, 
\label{eq:A14} \\ 
\beta &=& \frac{\hbar c q}{E+m c^2-V_b+S_b},
\label{eq:A15} \\
\gamma &=& \frac{\hbar c p}{E+m c^2-V_a+S_a},
\label{eq:A16}
\end{eqnarray}
where $\sqrt{z}$ signifies the square root of a complex $z$:
for a real $z$, $\sqrt{z}=z^{1/2}\:\Theta(z)+ i \:(-z)^{1/2}\:\Theta(-z)$
with the Heaviside function $\Theta(z)$.

Matching conditions of wave functions at boundaries at $x=\pm b$ and $x=\pm a$ yield
\begin{eqnarray}
\left( {\begin{array}{*{20}c}
   {e^{ - iqb} } & {e^{iqb} }  \\
   {\beta \: e^{ - iqb} } & { - \beta \:e^{ iqb} }  \\
\end{array}} \right)\left( {\begin{array}{*{20}c}
   {A_1 }  \\
   {B_1 }  \\
\end{array}} \right) &=& \left( {\begin{array}{*{20}c}
   {e^{ - ikb} } & {e^{ikb} }  \\
   {\alpha \: e^{ - ikb} } & { - \alpha\: e^{ ikb} }  \\
\end{array}} \right)\left( {\begin{array}{*{20}c}
   {A_2 }  \\
   {B_2 }  \\
\end{array}} \right),
\label{eq:A17}
\end{eqnarray}

\begin{eqnarray}
\left( {\begin{array}{*{20}c}
   {e^{ -ika} } & {e^{ika} }  \\
   {\alpha\: e^{-ika} } & { - \alpha \:e^{ika} }  \\
\end{array}} \right)\left( {\begin{array}{*{20}c}
   {A_2 }  \\
   {B_2 }  \\
\end{array}} \right) &=& \left( {\begin{array}{*{20}c}
   {e^{ -ipa} } & {e^{ipa} }  \\
   {\gamma \:e^{ -ipa} } & { - \gamma \:e^{ipa} }  \\
\end{array}} \right)\left( {\begin{array}{*{20}c}
   {A_3 }  \\
   {B_3 }  \\
\end{array}} \right),
\label{eq:A18}
\end{eqnarray}

\begin{eqnarray}
\left( {\begin{array}{*{20}c}
   {e^{ ipa} } & {e^{-ipa} }  \\
   {\gamma \:e^{ ipa} } & { - \gamma \:e^{ - ipa} }  \\
\end{array}} \right)\left( {\begin{array}{*{20}c}
   {A_3 }  \\
   {B_3 }  \\
\end{array}} \right) &=& \left( {\begin{array}{*{20}c}
   {e^{ ika} } & {e^{-ika} }  \\
   {\alpha \: e^{ ika} } & { - \alpha \: e^{ - ika} }  \\
\end{array}} \right)\left( {\begin{array}{*{20}c}
   {A_4 }  \\
   {B_4 }  \\
\end{array}} \right),
\label{eq:A19}
\end{eqnarray}

\begin{figure}
\begin{center}
\includegraphics[keepaspectratio=true,width=150mm]{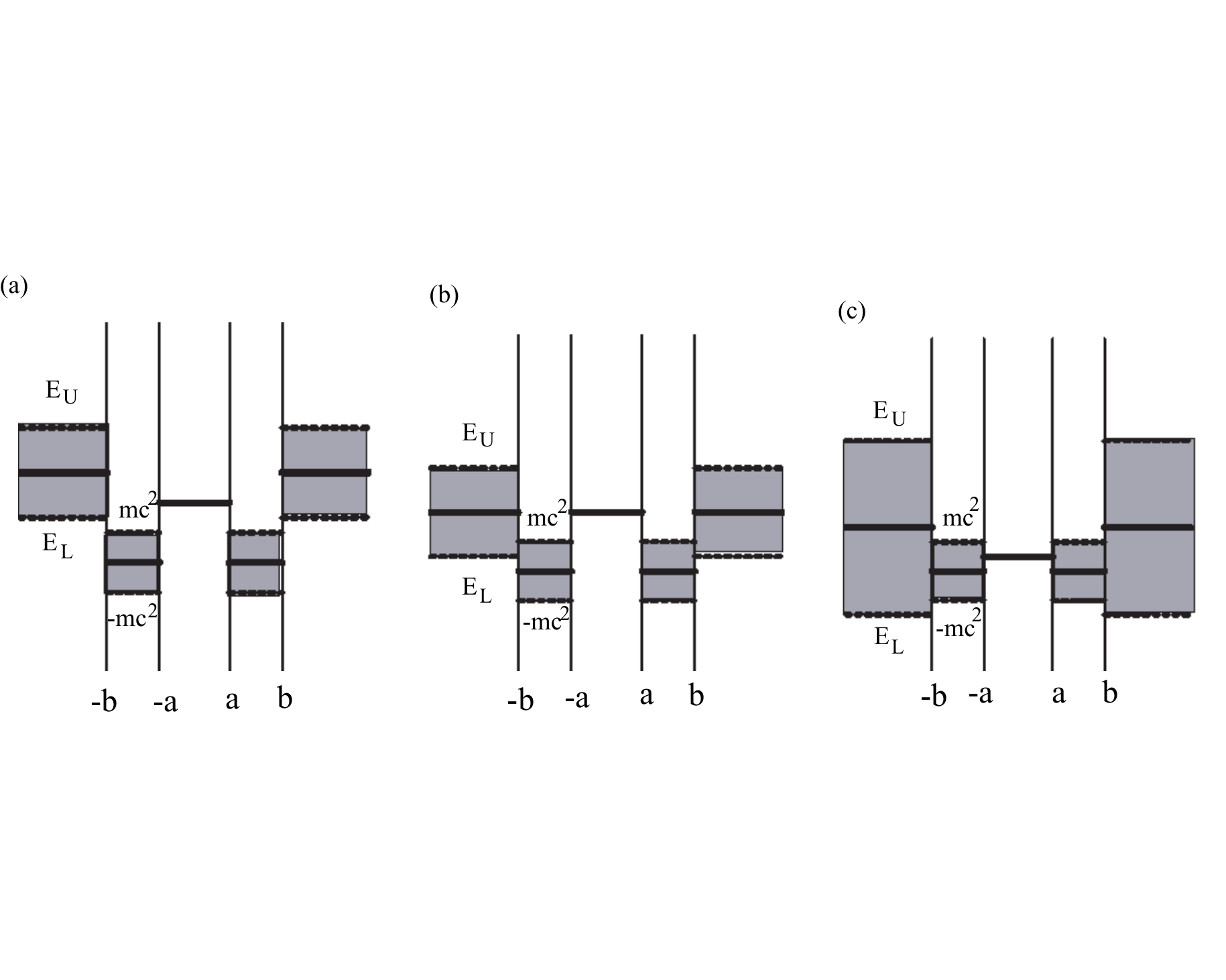}
\end{center}
\caption{
(Color online) 
Critical energy levels for (a) $V_b-S_b \geq 2 mc^2$, (b) $0 \leq V_b-S_b <2 mc^2$, 
and (c) $V_b -S_b <0$ ($E_U=V_b+S_b+mc^2$, $E_L=V_b-S_b-mc^2$),
wave vectors of $k$ and $p$ being purely imaginary in dark areas (see text).
}
\label{fig2}
\end{figure}

\begin{eqnarray}
\left( {\begin{array}{*{20}c}
   {e^{ ikb} } & {e^{-ikb} }  \\
   {\alpha\: e^{ ikb} } & { - \alpha\: e^{ - ikb} }  \\
\end{array}} \right)\left( {\begin{array}{*{20}c}
   {A_4 }  \\
   {B_4 }  \\
\end{array}} \right) &=& \left( {\begin{array}{*{20}c}
   {e^{ iqb} } & {e^{-iqb} }  \\
   {\beta\: e^{ ipb} } & { - \beta\: e^{ - ipb} }  \\
\end{array}} \right)\left( {\begin{array}{*{20}c}
   {A_5 }  \\
   {B_5 }  \\
\end{array}} \right).
\label{eq:A20}
\end{eqnarray}
By matrix calculation, we obtain 
\begin{eqnarray}
\left( {\begin{array}{*{20}c}
   {A_{i} }  \\
   {B_{i} }  \\
\end{array}} \right) &=& {\sf M}_{i \:i+1}
\left( {\begin{array}{*{20}c}
   {A_{i+1} }  \\
   {B_{i+1} }  \\
\end{array}} \right)
\hspace{1cm} \mbox{for $i=1-4$},
\label{eq:B1}
\end{eqnarray}
yielding
\begin{eqnarray}
\left( {\begin{array}{*{20}c}
   {A_1 }  \\
   {B_1 }  \\
\end{array}} \right) &=& {\sf T}
\left( {\begin{array}{*{20}c}
   {A_5 }  \\
   {B_5 }  \\
\end{array}} \right)
= \left( {\begin{array}{*{20}c}
   {T_{11} } & {T_{12} }  \\
   {T_{21} } & {T_{22} }  \\
\end{array}} \right)\left( {\begin{array}{*{20}c}
   {A_5 }  \\
   {B_5 }  \\
\end{array}} \right),
\label{eq:B2}
\end{eqnarray}
where the transfer matrix ${\sf T}$ given by 
${\sf T}={\sf M}_{1 \:2}\:{\sf M}_{2 \:3}\:{\sf M}_{3 \:4}\:{\sf M}_{4 \:5}$
includes information on the properties of a particle under consideration.

\subsection{Bound-state condition}
In order to obtain eigenvalues of a bounded particle, we set $A_1=B_5=0$ 
in Eq. (\ref{eq:B2}), 
which is satisfied by $T_{11}=0$.
After some matrix manipulations, we obtain the eigenvalue condition given by
\begin{eqnarray}
T_{11}&=&\frac{e^{2 i qb}}{16 \alpha^2 \beta \gamma}
\{
(\alpha+\gamma)^2 \left[(\alpha+\beta)^2 \:e^{2i[k(a-b)-pa]} 
- (\alpha-\beta)^2 \:e^{-2i[k(a-b)- pa]} \right] \nonumber \\
&+& (\alpha-\gamma)^2 \left[(\alpha-\beta)^2 \:e^{-2i[k(a-b)+pa]} 
- (\alpha+\beta)^2  \:e^{2i[k(a-b)+ pa]} \right] \nonumber \\
&+& 2(\alpha^2-\beta^2)(\alpha^2-\gamma^2)  
\left[\:e^{2i pa} - e^{-2ipa} \right]
\}=0.
\label{eq:B3}
\end{eqnarray}
Equation (\ref{eq:B3}) determines both even- and odd-parity solutions.

It is necessary to solve the transcendental complex equation given by Eq. (\ref{eq:B3})
in order to obtain eigenvalues of a bounded particle.
Once an eigenvalue $E=E_n$ for an index $n$ ($=1,2, \cdots $) is obtained, 
we may successively determine coefficients
of $A_i$ and $B_i$ ($i=2-4$) and $B_1$, starting from assumed coefficients of $A_5=C$
and $B_5=0$ by using Eq. (\ref{eq:B1}).  
The magnitude of the assumed $C$ is determined by the normalization condition for 
the density probability $\rho(x)$ given by
\begin{eqnarray}
\int_{-\infty}^{\infty} \rho(x)\:dx &=& 1,
\label{eq:B8}
\end{eqnarray}
with
\begin{eqnarray}
\rho(x) &=& \vert \psi_+(x) \vert^2 + \vert \psi_-(x) \vert^2,
\label{eq:B9}
\end{eqnarray}
which may be analytically evaluated.

\begin{figure}
\begin{center}
\includegraphics[keepaspectratio=true,width=100mm]{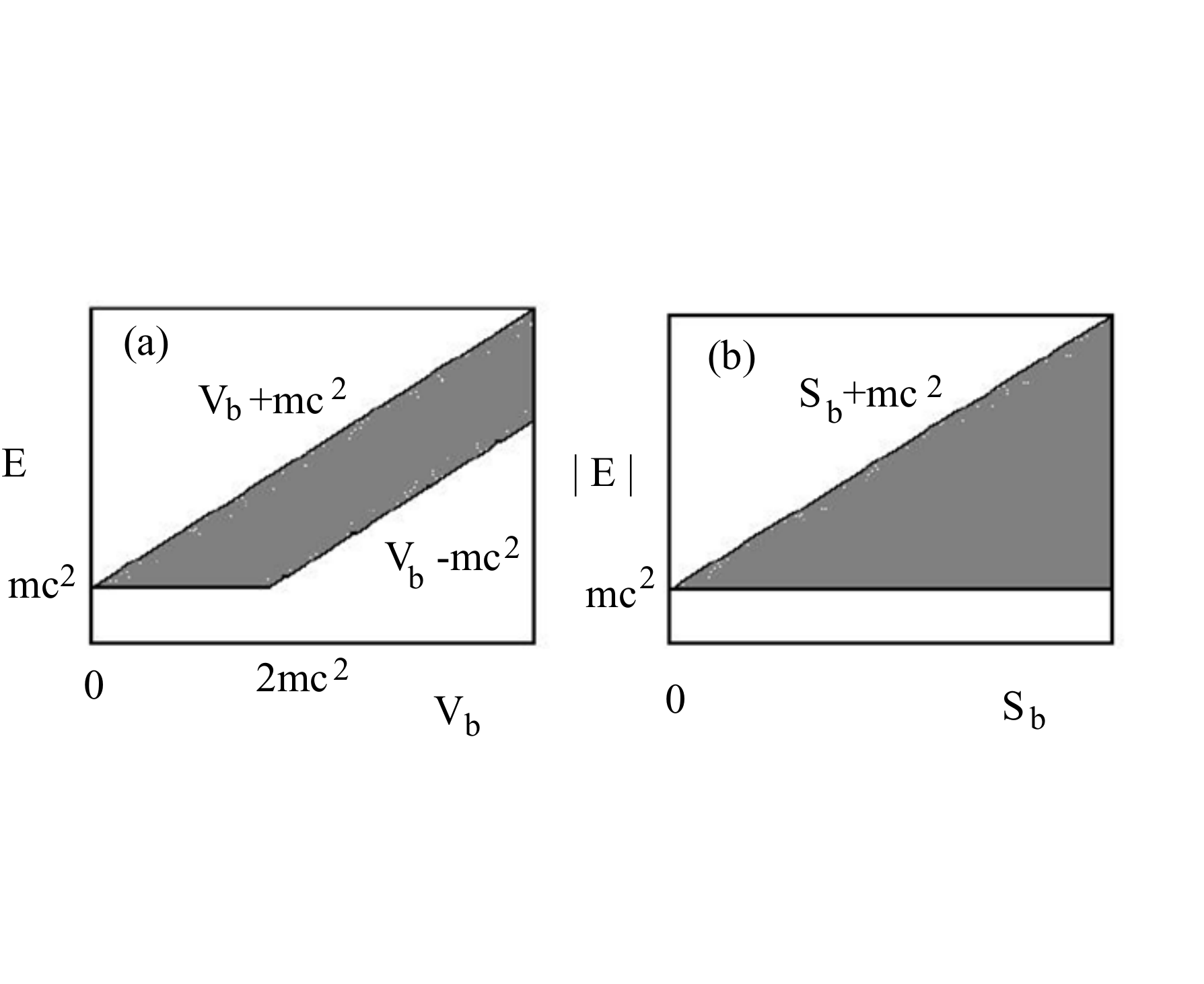}
\end{center}
\caption{
(Color online)
Energy ranges of bound states for (a) the VDSP [Eq. (\ref{eq:D4})] and 
(b) the SDSP [Eq. (\ref{eq:D5})], 
bound states existing in shaded regions.
The energy range for EDSP is expressed by (b) if we read
$S_b+mc^2 \rightarrow 2 S_b +mc^2$ [Eq. (\ref{eq:D6})]. 
}
\label{fig3}
\end{figure}

\subsection{Bound-state energy range}
We examine the energy range for bound states.
Depending on magnitudes of $V_b-S_b$ and $mc^2$,  
the properties of wave vectors $k$ and $q$ change in the three
cases: (A) $V_b-S_b \geq 2 m c^2$, (B) $0 \leq V_b-S_b < 2 m c^2$
and (C) $V_b-S_b < 0$, as shown in Figs. \ref{fig2}(a), \ref{fig2}(b) 
and \ref{fig2}(c), respectively, 
where $k$ and $p$ are purely imaginary in dark areas.
Bound states exist when $k$ is real but $q$ is purely imaginary
which lead to plane waves in regions II and IV
and evanescent waves in regions I and V.  
When the above condition is satisfied, bound states exist independently of 
whether the wave vector $p$ in the region III is real or imaginary.
We obtain such energy regions for bound states given by
\begin{eqnarray}
E_L &<& E < E_U \hspace{4.2cm}
\mbox{for case A ($V_b-S_b \geq 2 mc^2$)},
\label{eq:D1}\\
mc^2 &<& E < E_U \hspace{4.2cm}
\mbox{for case B ($0 \leq V_b-S_b < 2 mc^2 $)},
\label{eq:D2}\\
m c^2 &<& E < E_U \hspace{0.2cm} {\rm or} \hspace{0.2cm}E_L < E < -mc^2 
\hspace{0.5cm}\mbox{for case C ($V_b-S_b < 0$)},
\label{eq:D3}
\end{eqnarray}
where
\begin{eqnarray}
E_U &=& V_b+S_b+m c^2, \\
E_L &=& V_b-S_b-m c^2.
\end{eqnarray}
In the so-called Klein region: $mc^2 < E< V_b-mc^2$ with $S_b=0$ in the case A,
we obtain oscillating waves in regions I and V, and then no bound states are realized
[see Fig. \ref{fig2}(a)]. 
The negative $E$ of $V_b-S_b-mc^2 < E < -mc^2$ in Eq. (\ref{eq:D3})
expresses the bound state for an antiparticle.
In the special cases of (1) $S_b=0$ (VDSP), (2) $V_b=0$ (SDSP), 
and (3) $S_b=V_b$ (EDSP), the bound-state condition becomes
\begin{eqnarray}
{\rm max}(V_b-mc^2, mc^2) &<& E < V_b+mc^2 
\hspace{0.5cm}\mbox{for $S_b=0$ (VDSP)}, 
\label{eq:D4} \\
mc^2 &<& \vert E \vert < S_b+mc^2 
\hspace{0.5cm}\mbox{for $V_b=0$ (SDSP)}, 
\label{eq:D5} \\
mc^2 &<& E < 2 S_b+mc^2 \hspace{0.5cm} 
\mbox{for $S_b=V_b$ (EDSP)}.
\label{eq:D6}
\end{eqnarray}
Bound-state energy ranges given by Eqs. (\ref{eq:D4}) and (\ref{eq:D5}) 
for the VDSP and SDSP 
are shown in Figs. \ref{fig3}(a) and  \ref{fig3}(b), respectively,
where bound states exist in shaded regions.
The bound-state range for EDSP is expressed by Fig. \ref{fig3}(b)
where $S_b+mc^2$ is replaced by $2 S_b +mc^2$ [Eq. (\ref{eq:D6})].
We note that the energy range for the VDSP in Fig. \ref{fig3}(a)
is quite different from those for the SDSP and EDSP in Fig. \ref{fig3}(b).
The bound-state region given by Eqs. (\ref{eq:D1})-(\ref{eq:D3})
which is derived by physical consideration, has been numerically confirmed by 
the eigenvalue condition given by Eq. (\ref{eq:B3}).

\subsection{Single square-well limit}
In the limit of $V_a=0$ and $S_a=0$, 
or in the limit of $a=0$ where the double square-well potential reduces to 
the single square-well potential, Eq. (\ref{eq:B3})
becomes
\begin{eqnarray}
T_{11} &=& \frac{e^{2 i qb}}{4 \alpha \beta} \left[ (\alpha+\beta)^2 \:e^{-2 ikb}
- (\alpha-\beta)^2 \:e ^{2ikb}\right] = 0,
\end{eqnarray}
leading to
\begin{eqnarray}
\tan(2 k b) &=& \frac{2 \alpha \kappa}{\alpha^2-\kappa^2}
\hspace{0.5cm}\left( \kappa=\sqrt{\frac{mc^2+V_b+S_b-E}{mc^2-V_b+S_b+E}} \right).
\label{eq:B10}
\end{eqnarray}
For the vector potential only ($S_b=0$), Eq. (\ref{eq:B10}) becomes
\begin{eqnarray}
\tan(2 k b) &=& \frac{2 \alpha \kappa_v}{\alpha^2-\kappa_v^2} 
\hspace{1cm}\left( \kappa_v=\sqrt{\frac{mc^2+V_b-E}{mc^2-V_b+E} } \right),
\label{eq:B10b}
\end{eqnarray}
which denotes the condition for the single vector square-well potential \cite{Greiner81,Coulter71}.
On the other hand, for the scalar potential only ($V_b=0$), Eq. (\ref{eq:B10}) becomes
\begin{eqnarray}
\tan(2 k b) &=& \frac{2 \alpha \kappa_s}{\alpha^2-\kappa_s^2} 
\hspace{1cm}\left( \kappa_s=\sqrt{\frac{mc^2+S_b-E}{mc^2+S_b+E} } \right),
\label{eq:B10c}
\end{eqnarray}
which expresses the condition for the single scalar square-well potential.
In particular in the limit of infinite confining potential with $S_b \rightarrow \infty$,
Eq. (\ref{eq:B10c}) yields \cite{Alberto96}
\begin{eqnarray}
\tan(2 k b) &=& - \frac{\hbar k}{mc}.
\label{eq:B10d}
\end{eqnarray}
Unfortunately such a limit of $V_b \rightarrow \infty$ cannot be taken for 
the vector single square-well potential 
in Eq. (\ref{eq:B10b}).

\subsection{Nonrelativistic limit}
Before going to model calculations, we examine the nonrelativistic limit of 
the bound-state condition in the Dirac equation 
with a shifted energy $E_s$ defined by 
\begin{eqnarray}
E_s=E-mc^2. 
\end{eqnarray}
In the nonrelativistic limit of $mc^2 \rightarrow \infty$, 
Eqs. (\ref{eq:A11})-(\ref{eq:A16}) become
\begin{eqnarray}
k &\rightarrow& \frac{\sqrt{2m E_s}}{\hbar}, \;\;\;
q \rightarrow \frac{\sqrt{2m (E_s-V_b-S_b)}}{\hbar}, \;\;\;
p \rightarrow \frac{\sqrt{2m (E_s-V_a-S_a)}}{\hbar}, \\
%
\alpha &\rightarrow& \frac{\hbar \:k}{2m c} , \;\;\;
\beta \rightarrow \frac{\hbar \:q}{2mc}, \;\;\;
\gamma \rightarrow \frac{\hbar \:p}{2mc},
\end{eqnarray}
with which the bound-state condition given by Eq. (\ref{eq:B3}) reduces to Eq. (\ref{eq:X6})
with Eqs. (\ref{eq:X1})-(\ref{eq:X3}) in the Schr\"{o}dinger equation,
if we read $E_s \rightarrow E$, $V_a+S_a \rightarrow V_a$ and $V_b+S_b \rightarrow V_b$. 
Equations (\ref{eq:D1})-(\ref{eq:D3}) become
\begin{eqnarray}
V_b+S_b &>& E_s > {\rm max}(V_b-S_b-2 m c^2, 0) \rightarrow 0.
\label{eq:C2}
\end{eqnarray}
Then the bound-state condition of the Dirac equation given by 
Eqs. (\ref{eq:B3}) and (\ref{eq:C2}) 
in the nonrelativistic limit is equivalent to that of the Schr\"{o}dinger equation 
given by Eqs. (\ref{eq:X6}) and (\ref{eq:X7}).

\section{Model calculations}
The transcendental complex equation (\ref{eq:B3}) has been solved with the use of MATHEMATICA.
We will separately present model calculations for (1) VDSP, (2) SDSP and (3) EDSP 
in Secs. III A, III B and III C, respectively,
adopting atomic units of $m=\hbar=e=1$ and then $c=137.036$. 

\subsection{Vector potential only ($S(x)=0$)}
\begin{figure}
\begin{center}
\includegraphics[keepaspectratio=true,width=80mm]{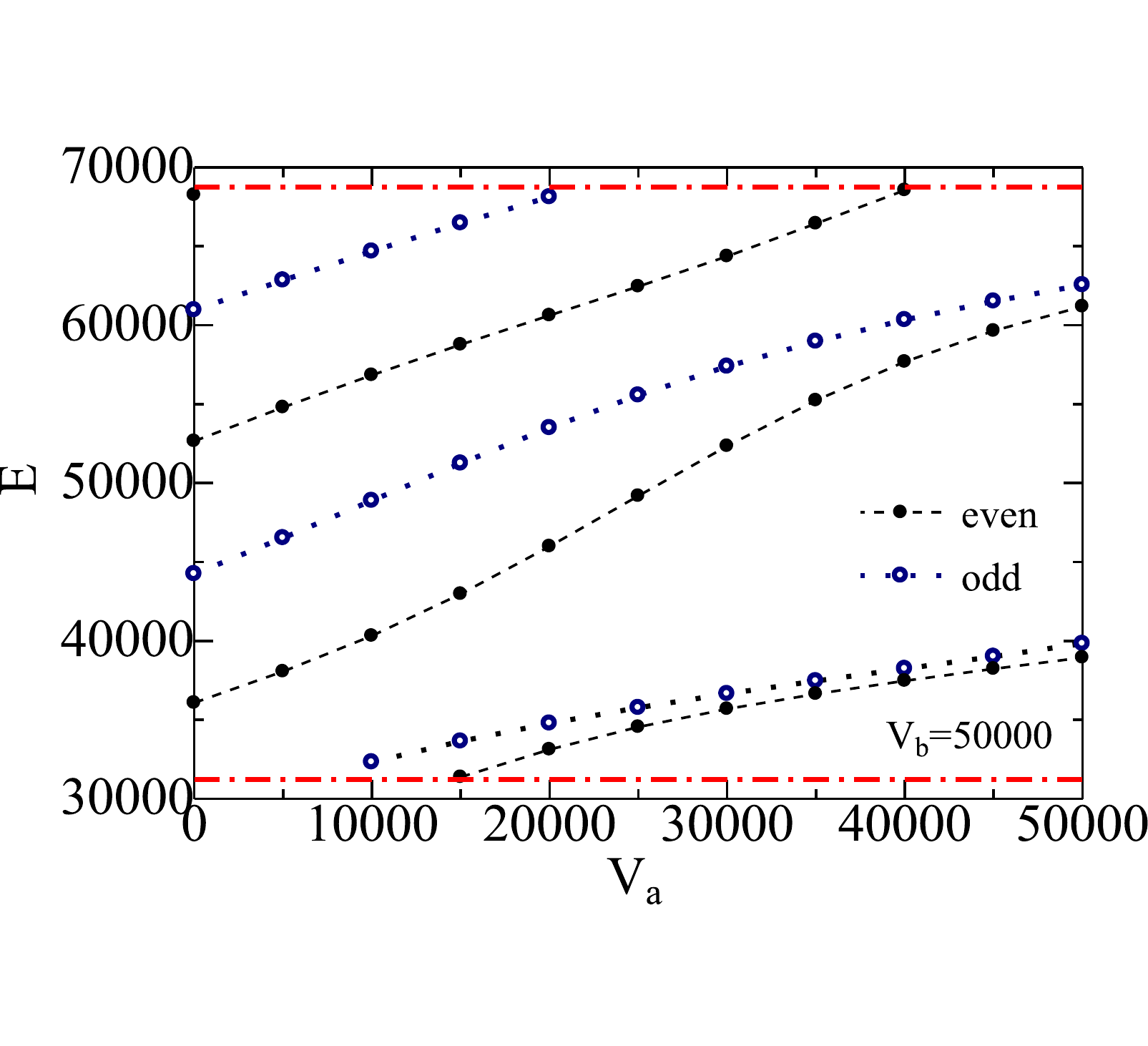}
\end{center}
\caption{
(Color online) 
The $V_a$ dependence of eigenvalues $E_n$ for the VDSP
($V_b=50000$, $S_a=S_b=0$, $a=0.01$ and $b=0.02$):
bound states appear between lower and upper limits expressed by chain curves.
For eigenstates depicted by filled and open circles,
$\psi_+(x)$ has even- and odd-parity wave functions, respectively,
while $\psi_{-}(x)$ has the opposite parity (see text).
}
\label{fig4}
\end{figure}

First we consider the case of the VDSP,
changing $V_a$ with fixed $V_b=50000$, $S_a=S_b=0$, $a=0.01$ and $b=0.02$.
Calculated eigenvalues are plotted as a function of $V_a$ in Fig. \ref{fig4},
numerical values of some eigenvalues being shown also in Table 1. 
Filled and open circles denote eigenvalues for which $\psi_+(x)$ has the even and
odd parities, respectively, whereas $\psi_-(x)$ has the opposite parity.
The number of eigenvalues for $(V_a,V_b)=(0,50000)$ 
in the range of $31231 < E_n < 68769$ is five ($n=1-5$).
With increasing $V_a$, eigenvalues are gradually increased.
For $V_a > 10000$, new eigenstates appear at $E_1 \gtrsim 32000$.
With furthermore increasing $V_a$, quasi-degenerate pair states appear:
for $V_a=50000$, $E_1 \simeq E_2$.  and $E_3 \simeq E_4$.

\begin{center}
\begin{tabular}[b]{|c|c|c|c|c|c|c|}
\hline
$V_a$ & $n=1$ & $n=2$ & $n=3$ &  $n=4$ & $n=5$ & $n=6$  \\ 
\hline \hline
$\;\;0\;\;$ & 36085  & 44246 & 52660 & 60964 & 68254 & $-$ \\
10000 &  32325  &  40323 & 48890 & 56839 & 64679  &  $-$  \\
20000 & 33124 & 34783 & 45988  & 53502 & 60624  & 68125  \\
30000 & 35693 & 36655  &  52348  & 57389  & 64365 &   $-$  \\
40000 & 37475 &  38240 & 57687  & 60339 & 68554 &  $-$  \\
50000 & 38948  & 39831 & 61189  & 62552 &  $-$ &  $-$  \\
\hline
\end{tabular}
\end{center}
{\it Table 1} Eigenvalues $E_n$ as a function of $V_a$  
for the VDSP with $V_b=50000$, $S_a=S_b=0$, $a=0.01$ and $b=0.02$,
the index $n$ being assigned from the lowest eigenvalue (see Fig. \ref{fig4}).

\begin{figure}
\begin{center}
\includegraphics[keepaspectratio=true,width=150mm]{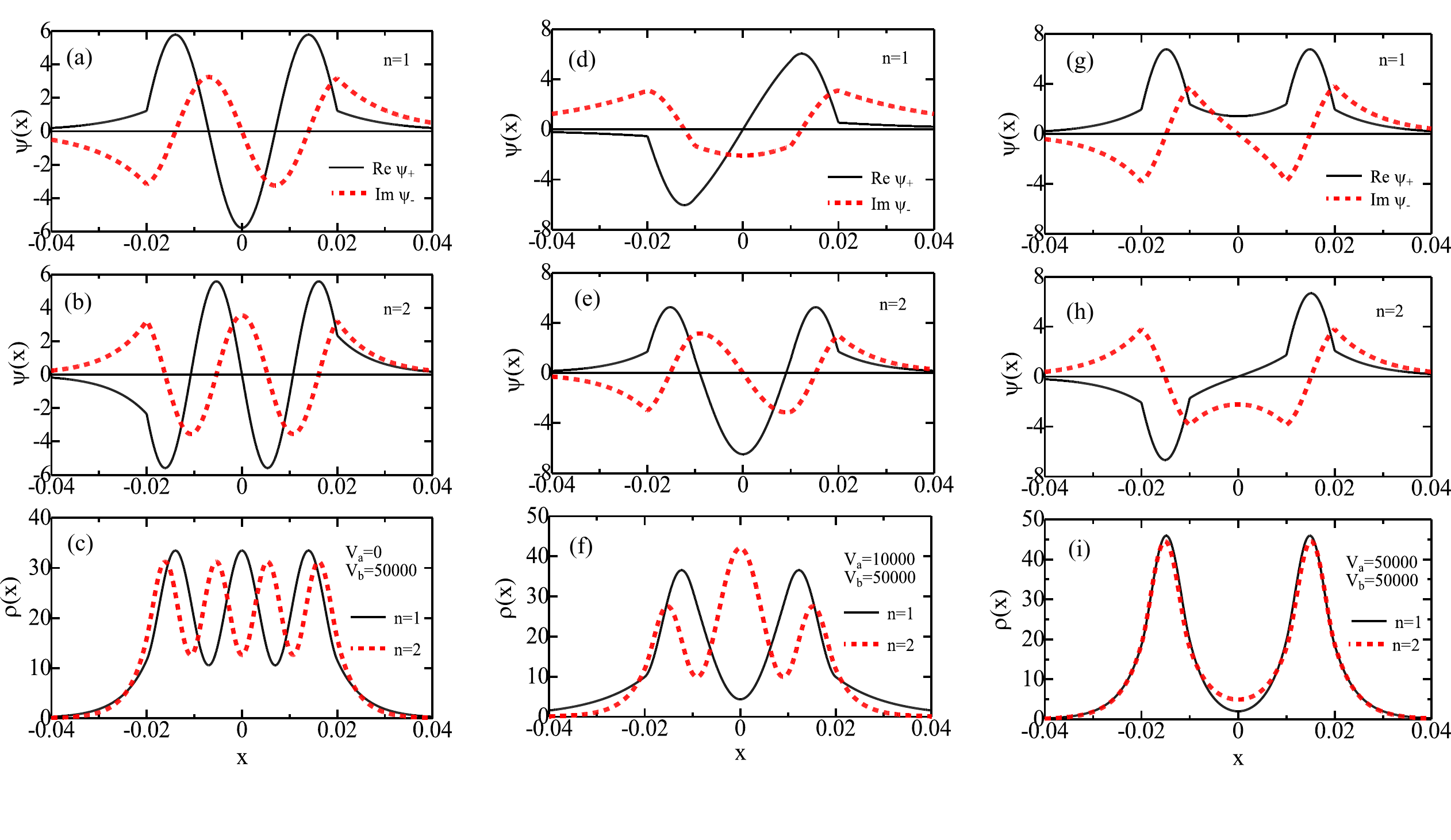}
\end{center}
\caption{
(Color online)
Wave function $\psi_{\pm}(x)$ and density probability $\rho(x)$ for the VDSP; 
$\psi_{\pm}(x)$ for (a) $n=1$ and (b) $n=2$, and 
(c) $\rho(x)$ for $n=1$ and $n=2$
with $(V_a,V_b)=(0, 50000)$;
$\psi_{\pm}(x)$ for (d) $n=1$ and (e) $n=2$, and 
(f) $\rho(x)$ for $n=1$ and $n=2$ 
with $(V_a,V_b)=(10000,50000)$;
$\psi_{\pm}(x)$ for (g) $n=1$ and (h) $n=2$, and 
(i) $\rho(x)$ for $n=1$ and $n=2$ 
with $(V_a,V_b)=(50000,50000)$.
In (a), (b), (d), (e), (g) and (h), solid and dashed curves express
Re $\psi_+(x)$ and Im $\psi_-(x)$, respectively [Im $\psi_+(x)=$ Re $\psi_-(x)=0$].
In (c), (f) and (i), solid and dashed curves denote $\rho(x)$ for $n=1$ and $n=2$, 
respectively ($a=0.01$ and $0.02$). 
}
\label{fig5}
\end{figure}

Figures \ref{fig5}(a) and \ref{fig5}(b) show
wave functions for $n=1$ and $n=2$, respectively, with $(V_a,V_b)=(0, 50000)$.
We note that Re $\psi_+(x)$ for $n=1$ with $V_b=50000$ in Figs. \ref{fig5}(a)
have three nodes in contrast with the conventional wisdom that the ground-state
wavefunction has a single node.
It is the case also for $n=2$ where 
number of nodes of Re $\psi_+(x)$ in Fig. \ref{fig5}(b) is four.
This is because wave vectors $k$ for
$E_1=38948$ and $E_2=39831$ with $V_b=50000$ have large values.
Figure \ref{fig5}(c) shows that density probabilities $\rho(x)$ for $n=1$ and $n=2$ 
have three and four peaks, respectively.

We introduce $V_a=10000$ in the central square potential, for which
the wave vector $p$ in the region III is real.
Solid (dashed) curves in Figs. \ref{fig5}(d) and \ref{fig5}(e)
show Re $\psi_+(x)$ (Im $\psi_-(x)$) for $n=1$ and $n=2$, respectively.
It is noted that the parity of $\psi_+(x)$ for $n=2$ is even while that for $n=1$ is odd.
This is because $\psi_+(x)$ for $n=2$ with
$V_a=10000$ has the same even parity as that for $n=1$ with $V_a=0$, 
as shown in Fig. \ref{fig4}.
Density probabilities $\rho(x)$ for $n=1$ and $n=2$ have two and three peaks, respectively,
in Fig. \ref{fig5}(f).

The value of $V_a$ is furthermore increased to $V_a=50000$, 
for which $p$ in the region III becomes imaginary.
Figures \ref{fig5}(g) and \ref{fig5}(h) show that
magnitudes of wave functions for $n=1$ and $n=2$ in the region III 
are much reduced compared to those in regions II and IV.
Then magnitudes of density probabilities in the region III become significantly smaller
than those in regions II and IV, as shown in Fig. \ref{fig5}(i). 

\subsection{Scalar potential only ($V(x)=0$)}
\begin{figure}
\begin{center}
\includegraphics[keepaspectratio=true,width=80mm]{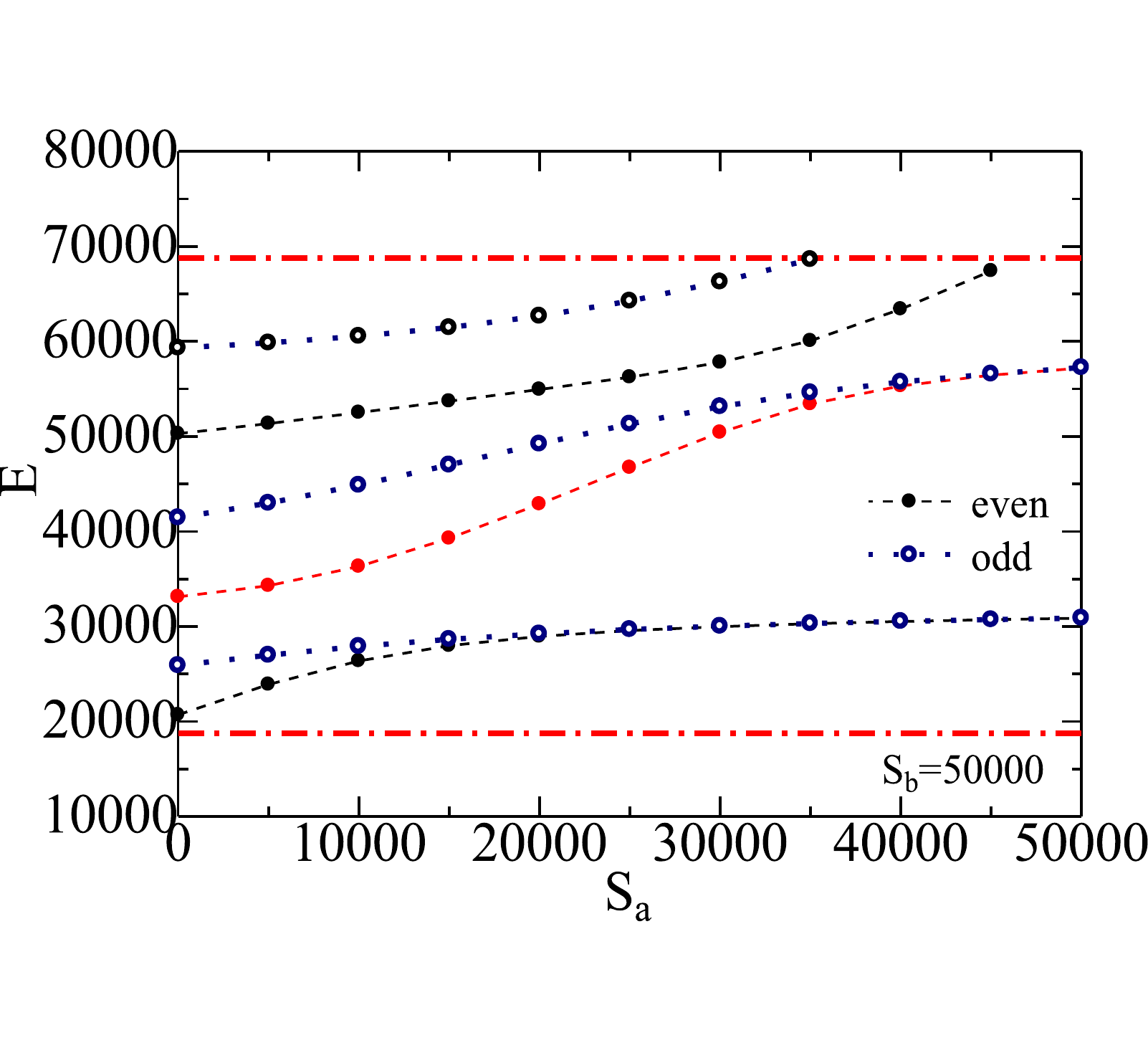}
\end{center}
\caption{
(Color online) 
The $S_a$ dependence of eigenvalues $E_n$ for the SDSP 
($S_b=10000$, $V_a=V_b=0$, $a=0.01$ and $b=0.02$):
bound states appear between lower and upper limits expressed by chain curves.
For eigenstates depicted by filled and open circles,
$\psi_{+}(x)$ has even- and odd-parity wave functions, respectively,
while $\psi_{-}(x)$ has the opposite parity.
}
\label{fig6}
\end{figure}

Next we study the case of the SDSP, changing
$S_a$ with fixed $S_b=50000$, $V_a=V_b=0$, $a=0.01$ and $b=0.02$.
Figure \ref{fig6} shows calculated eigenvalues as a function of $S_a$,
numerical figures of some results being shown also in Table 2.
Note that eigenvalues are given for a pair of $\pm E$ [Eq. (\ref{eq:D3})]
although we will hereafter consider the positive eigenvalue only.
For eigenvalues shown by filled and open circles, $\psi_+(x)$ ($\psi_-(x)$) has
the even and odd (odd and even) parities, respectively.
For $S_a=0$, we have six bound states within the allowed range 
of $ 18769 < E_n < 68769$ ($n=1-6$) between the lower and upper limits shown 
by dashed curves. With increasing $V_a$, eigenvalues are gradually increased.
For $S_a \geq 20000$, eigenvalues of $E_1$ and $E_2$ are quasi-degenerate,
but not degenerate \cite{Coutinho88}.
This trend is the same as that for the VDSP shown in Fig. \ref{fig4}.

\begin{center}
\begin{tabular}[b]{|c|c|c|c|c|c|c|}
\hline
$S_a$ & $n=1$ & $n=2$ & $n=3$ &  $n=4$ & $n=5$ & $n=6$  \\ 
\hline \hline
$\;\;0\;\;$ &  20708 & 25901 &  33130 & 41443 & 50290 & 59317 \\
10000 & 26381 &  27912 & 36321 & 44871 & 52510  & 60547   \\
20000 & 28963 & 29238 & 42876  & 49214 & 54937  & 62666  \\
30000 &  29994 & 30042 & 50417  & 53138 & 57791 &  66274 \\
40000 &  30539 & 30548 & 55317  & 55730 & 63393 &  $-$  \\
50000 & 30887  & 30888 & 57204 & 57251 &  $-$ &  $-$  \\
\hline
\end{tabular}
\end{center}
{\it Table 2} Eigenvalues $E_n$ as a function of $S_a$  
for the SDSP with $S_b=50000$, $V_a=V_b=0$, $a=0.01$ and $b=0.02$,
the index $n$ being assigned from the lowest eigenvalue (see Fig. \ref{fig6}).

\begin{figure}
\begin{center}
\includegraphics[keepaspectratio=true,width=150mm]{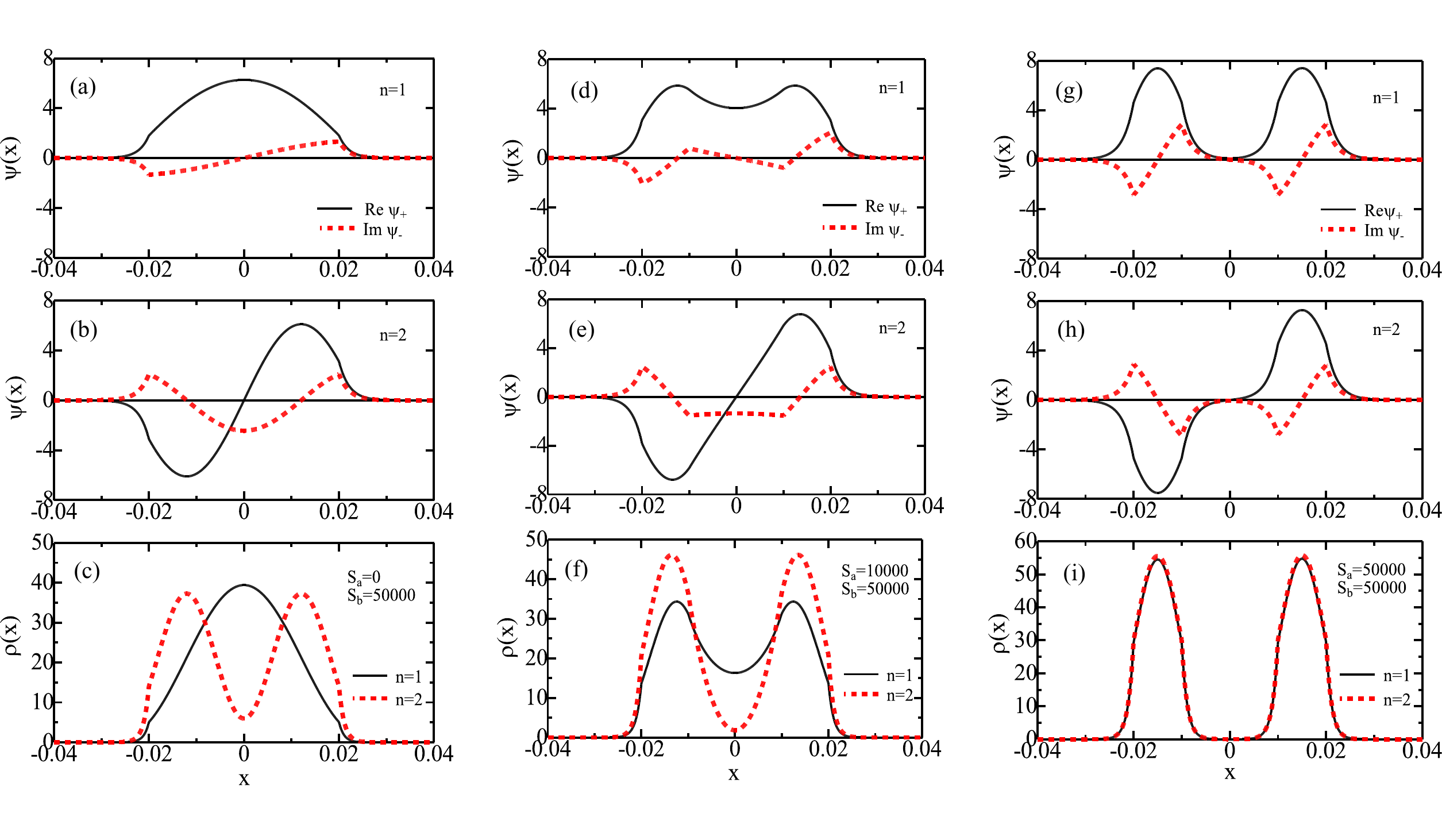}
\end{center}
\caption{
(Color online)
Wave function $\psi_{\pm}(x)$ and density probability $\rho(x)$ for SDSP; 
$\psi_{\pm}(x)$ for (a) $n=1$ and (b) $n=2$, and (c) $\rho(x)$ for $n=1$ and $n=2$
with $(S_a,S_b)=(0, 50000)$;
$\psi_{\pm}(x)$ for (d) $n=1$ and (e) $n=2$, and 
(f) $\rho(x)$ for $n=1$ and $n=2$ 
with $(S_a, S_b)=(10000, 50000)$;
$\psi_{\pm}(x)$ for (g) $n=1$ and (h) $n=2$, and 
(i) $\rho(x)$ for $n=1$ and $n=2$ 
with $(S_a, S_b)=(50000, 50000)$.
In (a), (b), (d), (e), (g) and (h), solid and dashed curves express
Re $\psi_+(x)$ and Im $\psi_-(x)$, respectively [Im $\psi_+(x)=$ Re $\psi_-(x)=0$].
In (c), (f) and (i), solid and dashed curves denote $\rho(x)$ for $n=1$ and $n=2$, 
respectively ($a=0.01$ and $0.02$). 
}
\label{fig7}
\end{figure}

Calculated wavefunctions and density probabilities are plotted 
in Figs. \ref{fig7}(a)-\ref{fig7}(i).
Figures \ref{fig7}(a) and \ref{fig7}(b) show wavefunctions for $n=1$ and $n=2$,
respectively, for $(S_a,S_b)=(0, 50000)$, and Fig. \ref{fig7}(c) denotes 
relevant density probabilities.
With the central barrier potential of $S_a=10000$ for which
the wave vector $q$ becomes imaginary,
magnitudes of the wavefunction and probability density for $n=1$
at $-0.01 < x < 0.01$ are decreased, as shown in Figs. \ref{fig7}(d)-\ref{fig7}(f).
Figures \ref{fig7}(g)-\ref{fig7}(i) show that for a larger $S_a=50000$,
magnitudes of $\rho(x)$ and $\psi_{\pm}(x)$ almost completely vanish 
at $-0.01 < x < 0.01$.

\subsection{Equal Scalar and vector potentials ($S(x)=V(x)$)}

\begin{figure}
\begin{center}
\includegraphics[keepaspectratio=true,width=80mm]{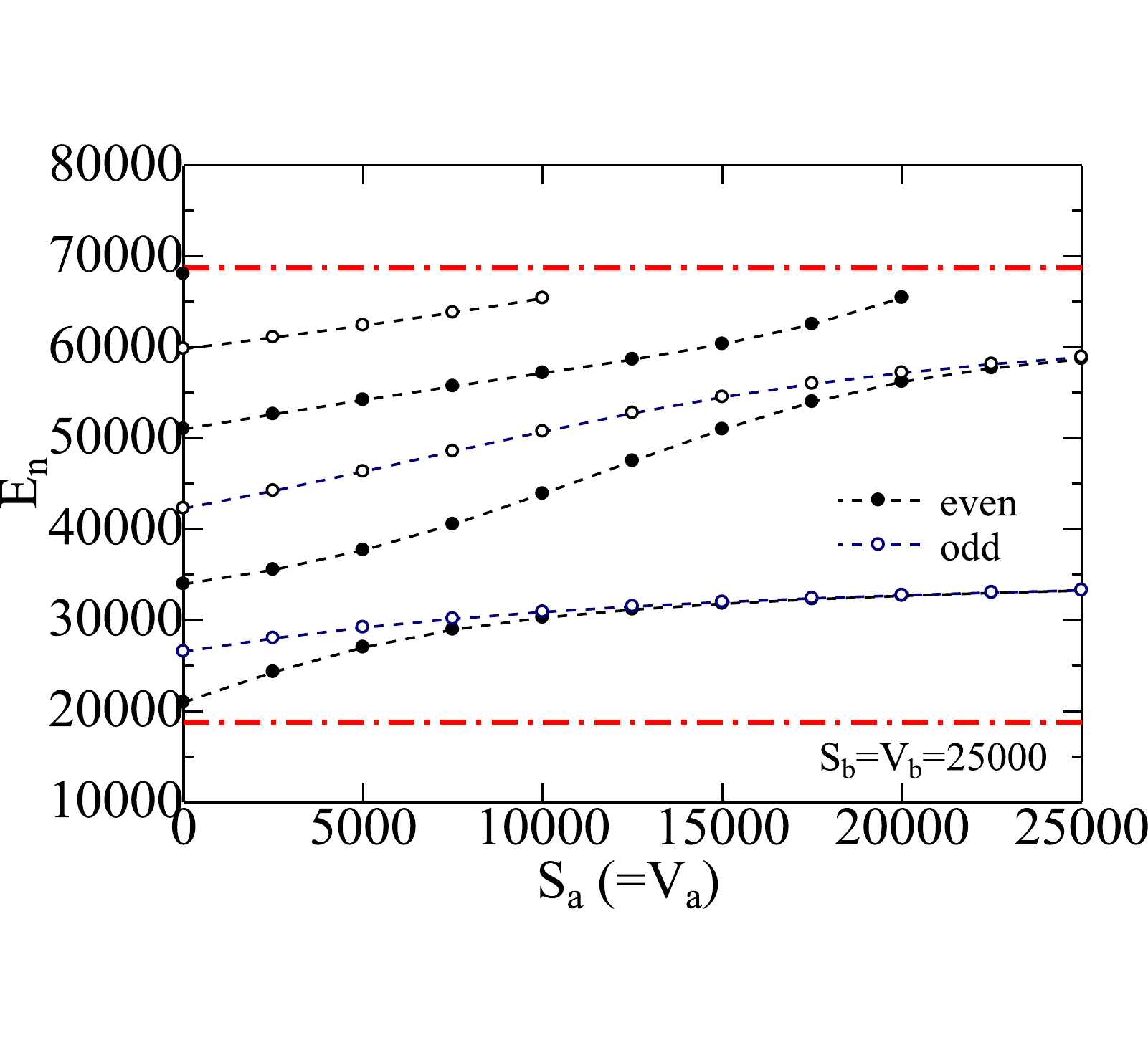}
\end{center}
\caption{
(Color online) 
The $S_a$ ($=V_a$) dependence of eigenvalues $E_n$ for EDSP
($S_a=V_a=25000$, $a=0.01$ and $b=0.02$):
bound states appear between lower and upper limits expressed by chain curves.
For eigenstates depicted by filled and open circles,
$\psi_{+}(x)$ has even- and odd-parity wave functions, respectively,
while $\psi_{-}(x)$ has the opposite parity.
}
\label{fig8}
\end{figure}

We study the case of the EDSP [$S(x)=V(x)$], 
for which the Dirac equation is expressed by one component equation given by
\begin{eqnarray}
\left[\hbar^2 c^2 \frac{d^2}{dx^2}+E^2-m^2c^4-2(mc^2+E)V(x) \right] \psi_+(x)=0, 
\end{eqnarray}
\begin{eqnarray}
\psi_-(x) = \left( \frac{-i\hbar c}{mc^2+E} \right) \frac{d}{dx} \psi_+(x). 
\end{eqnarray}
Figure \ref{fig8} shows eigenvalues calculated as a function of $S_a$ ($=V_a$)
for fixed $S_b=V_b=25000$, $a=0.01$ and $b=0.02$,
numerical values of some results being shown in Table 3.
We notice that the $V_a$ dependence of eigenvalues in Fig. \ref{fig8} is similar to
that for the SDSP shown in Fig. \ref{fig6}.

\begin{center}
\begin{tabular}[b]{|c|c|c|c|c|c|c|c|}
\hline
$S_a(=V_a)$ & $n=1$ & $n=2$ & $n=3$ &  $n=4$ & $n=5$ & $n=6$ &  $n=7$ \\ 
\hline \hline
$\;\;0\;\;$ & 20954 & 26527 & 33944  & 422623 & 50995 & 59814 & 68033 \\
5000 & 26990  &  29182  & 37672  & 46322 & 54189  & 62385  & $-$ \\
10000 & 30255 & 30882  & 43879  & 50724 & 57152  & 65381  & $-$ \\
15000 & 31791  & 31981 & 50973  & 54503 & 60328 &  $-$  & $-$\\
20000 &  32665 & 32731 &  56190  & 57167 & 65430 &  $-$  & $-$ \\
25000 & 33250  & 33276 & 58693 & 58910 &  $-$ &  $-$  & $-$ \\
\hline
\end{tabular}
\end{center}
{\it Table 3} Eigenvalues $E_n$ as a function of $S_a$ ($=V_a$)  
for the EDSP with $S_b=V_b=25000$, $a=0.01$ and $b=0.02$,
the index $n$ being assigned from the lowest eigenvalue (see Fig. \ref{fig8}).

\begin{figure}
\begin{center}
\includegraphics[keepaspectratio=true,width=150mm]{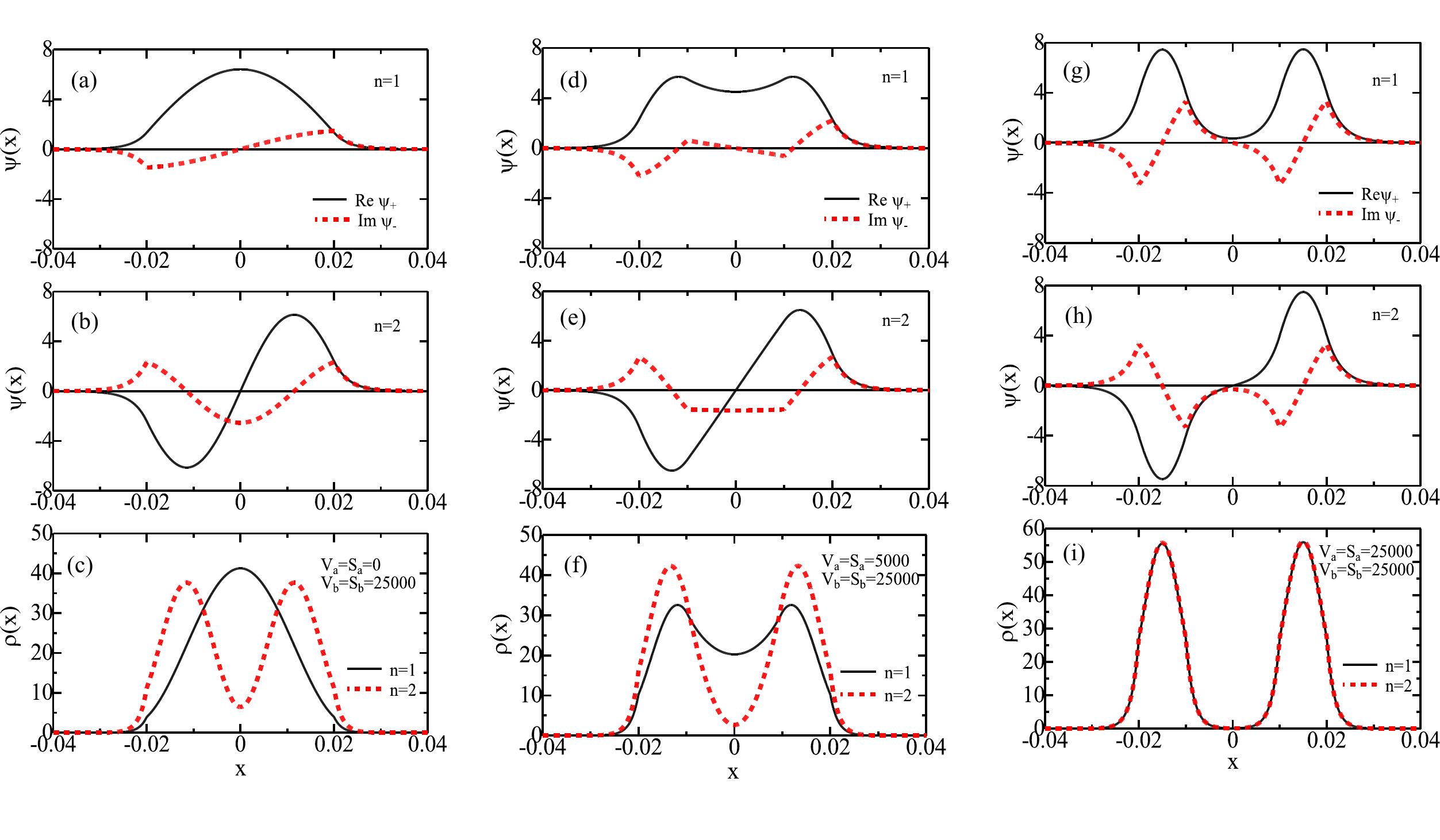}
\end{center}
\caption{
(Color online)
Wave function $\psi_{\pm}(x)$ and density probability $\rho(x)$ 
for the EDSP; 
$\psi_{\pm}(x)$ for (a) $n=1$ and (b) $n=2$, and (c) $\rho(x)$ for $n=1$ and $n=2$
with $(S_a,S_b)=(V_a,V_b)=(0, 25000)$;
$\psi_{\pm}(x)$ for (d) $n=1$ and (e) $n=2$, and 
(f) $\rho(x)$ for $n=1$ and $n=2$ 
with $(S_a,S_b)=(V_a,V_b)=(5000, 25000)$ ;
$\psi_{\pm}(x)$ for (g) $n=1$ and (h) $n=2$, and 
(i) $\rho(x)$ for $n=1$ and $n=2$ 
with  $(S_a,S_b)=(V_a,V_b)=(25000, 25000)$.
In (a), (b), (d), (e), (g) and (h), solid and dashed curves express
Re $\psi_+(x)$ and Im $\psi_-(x)$, respectively [Im $\psi_+(x)=$ Re $\psi_-(x)=0$].
In (c), (f) and (i), solid and dashed curves denote $\rho(x)$ for $n=1$ and $n=2$, 
respectively ($a=0.01$ and $0.02$). 
}
\label{fig9}
\end{figure}

Relevant wavefunctions and density probabilities are shown in 
Figs. \ref{fig9}(a)-\ref{fig9}(i). Although the wavevector $p$ is real
for the case of  $S_a=V_a=0$ in Figs. \ref{fig9}(a)-\ref{fig9}(c), 
it becomes imaginary for cases of $S_a=V_a=5000$ and $25000$
in Figs. \ref{fig9}(d)-\ref{fig9}(i).
Comparing Figs. \ref{fig9}(a)-\ref{fig9}(i) to Figs. \ref{fig7}(a)-\ref{fig7}(i),
we again notice that wavefunctions and probability densities
for the EDSP are quite similar to those for the SDSP.

\section{Discussion}
\subsection{Comparison with results of the Schr\"{o}dinger equation for the DSP}
We may apply the transfer-matrix method adopted in this study
to the Schr\"{o}dinger equation for the DSP.
A calculation for the Schr\"{o}dinger equation goes parallel to
that for the Dirac equation, details being provided in the Appendix.
The condition of bound states for the DSP is given by Eq. (\ref{eq:X6}),
which is ostensibly the same as Eq. (\ref{eq:B3}) 
if the relation: $\alpha/k=\beta/q=\gamma/p$ holds. 
%
From Eqs. (\ref{eq:D4})-(\ref{eq:D6}) and (\ref{eq:X7}),
the conceivable range of the bound-state energy $E$ ($> 0$) is given by
\begin{eqnarray}
{\rm max}(V_b-2 mc^2, 0) &<& E_s < V_b
\hspace{0.5cm} \mbox{for $S_b=0$ (VDSP)}, 
\label{eq:E1}\\
0 &<& E_s < S_b
\hspace{0.5cm} \mbox{for $V_b=0$ (SDSP)}, 
\label{eq:E2}\\
0 &<& E_s < 2 S_b
\hspace{0.5cm} \mbox{for $S_b=V_b$ (EDSP)}, 
\label{eq:E2b} \\
0 &<& E < V_b 
\hspace{2cm} \mbox{in the Schr\"{o}dinger equation},
\label{eq:E3}
\end{eqnarray}
where $E_s=E-mc^2$.
Bound-state ranges for SDSP and EDSP in the Dirac equation are similar 
to that in the Schr\"{o}dinger equation,
in contrast to that for the VDSP (Fig. \ref{fig3}).

\begin{figure}
\begin{center}
\includegraphics[keepaspectratio=true,width=70mm]{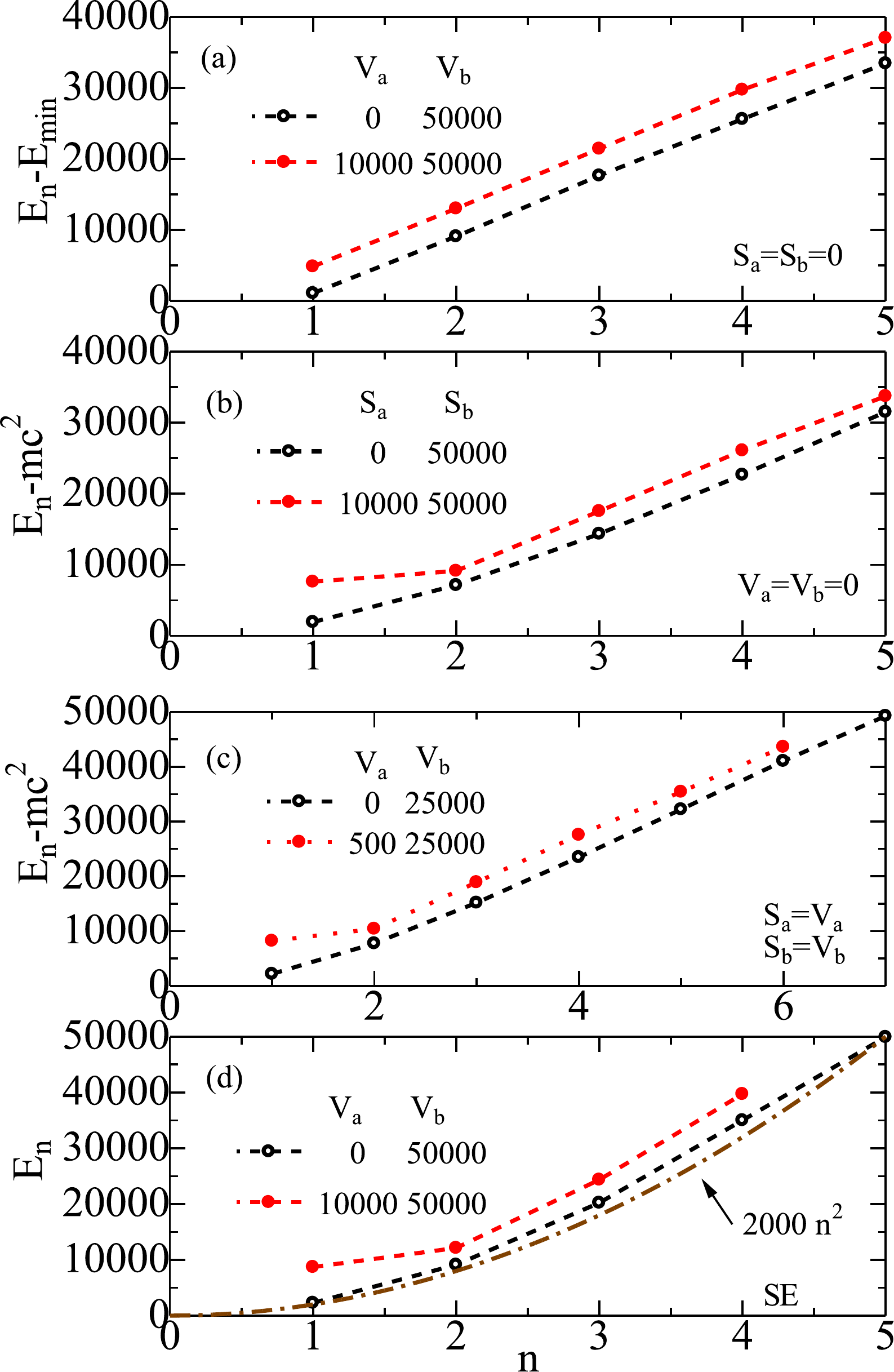}
\end{center}
\caption{
(a) The $n$ dependence of $E_n-E_{min}$ for (a) the VDSP of
$(V_a, V_b)=(0, 50000)$ (open circles) and $(10000, 50000)$ (filled circles)
with $(S_a, S_b)=(0, 0)$ and $E_{min}=V_b-mc^2=31231$;
(b) $E_n-mc^2$ for the SDSP of $(S_a, S_b)=(0, 50000)$ (open circles) 
and $(10000, 50000)$ (filled circles) with $(V_a, V_b)=(0, 0)$ and $mc^2=18769$;
(c) $E_n-mc^2$ with the EDSPs of $(V_a, V_b)=(S_a, S_b)=(0, 25000)$ (open circles) 
and $(5000, 25000)$ (filled circles); 
(d) $E_n$ of the Schr\"{o}dinger equation (SE)
for the DSP with $(V_a, V_b)=(0, 50000)$ (open circles) and $(10000, 50000)$ 
(filled circles), the chain curve denoting $2000 \:n^2$.
Dashed curves are plotted for guide of eye ($a=0.01$ and $b=0.02$).
}
\label{fig10}
\end{figure}

We have calculated eigenvalues of the Schr\"{o}dinger equation for the DSP. 
Calculated eigenvalues $E_n$ are plotted in Fig. \ref{fig10}(d) as a function of $n$
for two sets of $V_a$ and $V_b$.
Eigenvalues for $(V_a,V_b)=(0, 50000)$ and $(10000, 50000)$
approximately follow $E_n \simeq 2000 \:n^2$
which is shown by the chain curve. 
Note that the $E_n \propto n^2$ law is exactly realized in the limit of $V_b=\infty$ [Eq. (\ref{eq:X10})].

The $n$ dependence of $E-E_{min}$ for the VDSP studied in Secs. III A
is shown in Fig. \ref{fig10}(a) where $E_{min}={\rm max}(V_b-mc^2, mc^2)=31231$.
Figures \ref{fig10}(b) and \ref{fig10}(c)) show
the $n$-dependence of eigenvalues of $E-mc^2$ for the SDSP and EDSP, respectively, 
which are studied in Secs. III B and III C.
We expect from Figs. \ref{fig10}(a)-\ref{fig10}(c) that that the eigenvalue in the Dirac
equation for scalar and vector DSPs approximately follows a linear $n$ dependence 
for adopted parameters of $V_a \ll V_b$ and $S_a \ll S_b$.

\begin{figure}
\begin{center}
\includegraphics[keepaspectratio=true,width=150mm]{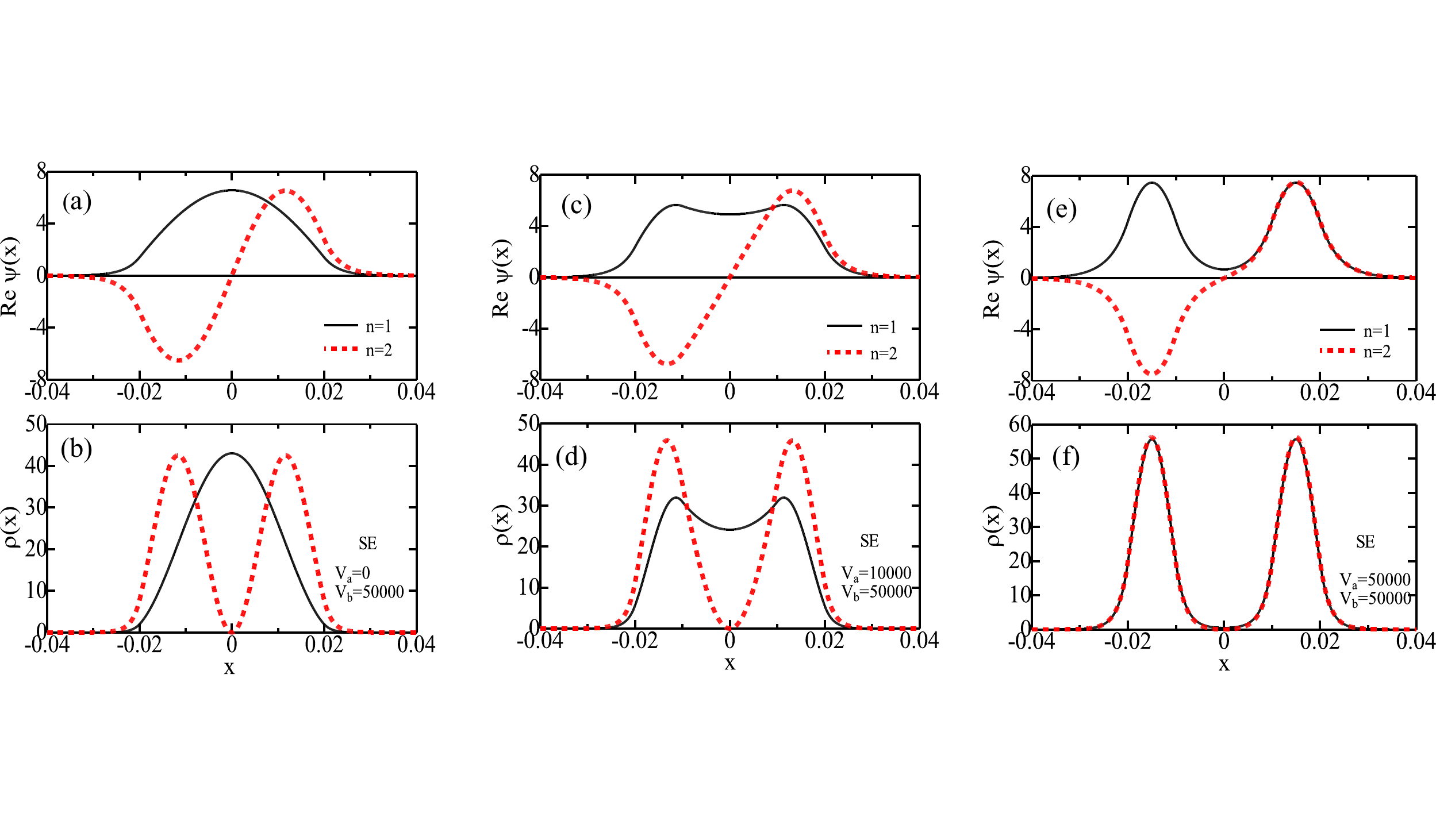}
\end{center}
\caption{
(Color online)
Wave function $\psi(x)$ and density probability $\rho(x)$ 
of the Schr\"{o}dinger equation (SE) for the DSP; 
(a) Re $\psi(x)$ and (b) $\rho(x)$ for $(V_a,V_b)=(0, 50000)$;
(c) Re $\psi(x)$ and (d) $\rho(x)$ for $(V_a,V_b)=(100000, 50000)$;
(e) Re $\psi(x)$ and (f) $\rho(x)$ for $(V_a,V_b)=(500000, 50000)$.
Solid and dashed curves express results for $n=1$ and $n=2$, respectively.
}
\label{fig11}
\end{figure}

We have calculated the wavefunction and probability density
in the Schr\"{o}dinger equation for the DSP with
$(V_a,V_b)=(0, 50000)$, $(10000, 50000)$ and $(50000, 50000)$ whose results 
are plotted in Figs. \ref{fig11}(a)-\ref{fig11}(f).
We note that wavefunctions and probability densities 
for $(V_a,V_b)=(0, 50000)$ in Figs. \ref{fig11}(a) and \ref{fig11}(b) 
are similar to $\psi_+(x)$ and $\rho(x)$ of the Dirac equation 
for the SDSP shown in Figs. \ref{fig7}(a)-\ref{fig7}(c)
and to those for the EDSP shown 
in Figs. \ref{fig9}(a)-\ref{fig9}(c), 
although they are quite different from those for the VDSP shown
in Figs. \ref{fig5}(a)-\ref{fig5}(c).
It is the case also for $(V_a,V_b)=(10000, 50000)$
in Figs. \ref{fig11}(c) and \ref{fig11}(d) 
and for $(V_a,V_b)=(50000, 50000)$
in Figs. \ref{fig11}(e) and \ref{fig11}(f). 

\begin{figure}
\begin{center}
\includegraphics[keepaspectratio=true,width=80mm]{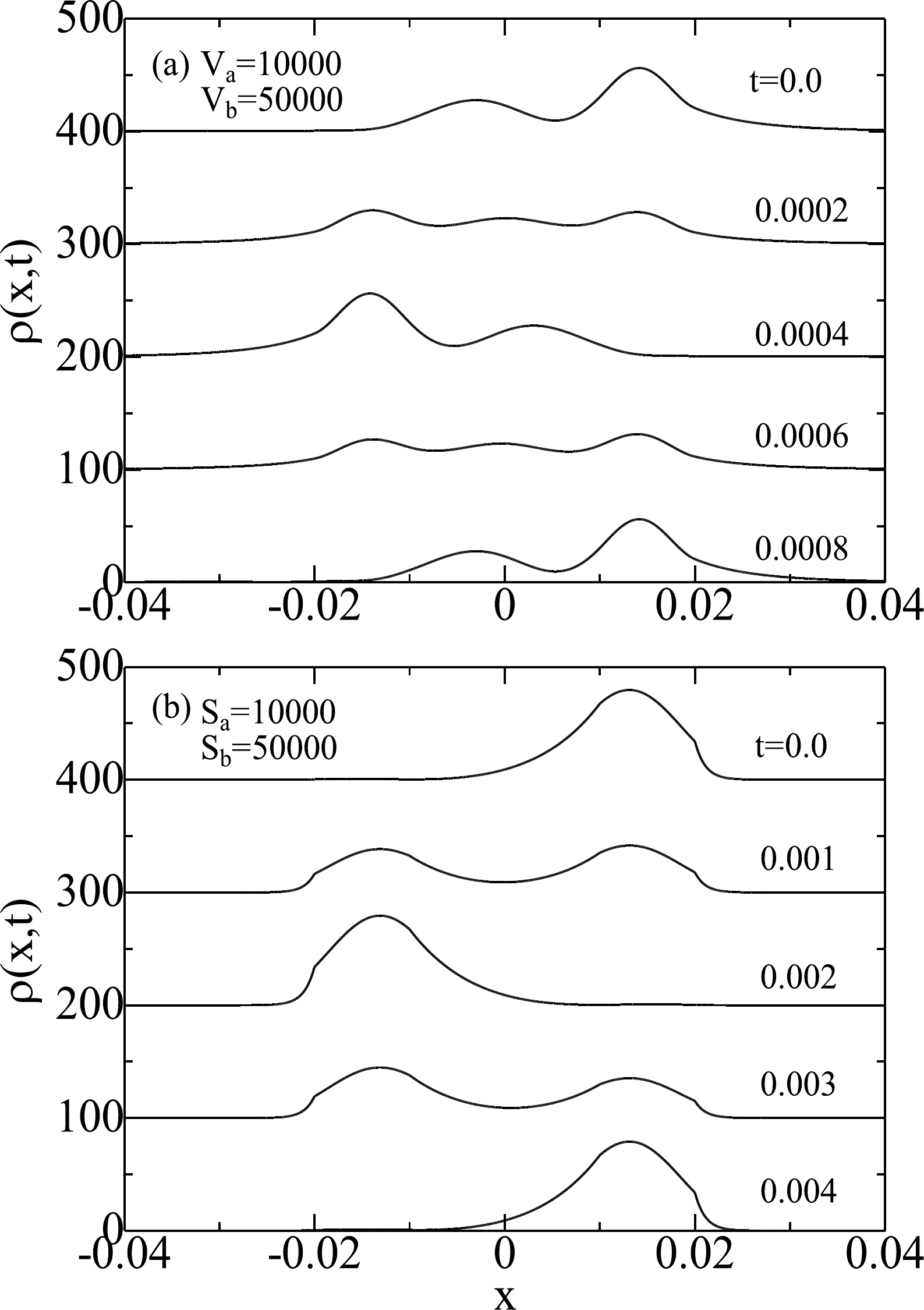}
\end{center}
\caption{
Time dependence of $\rho(x,t)$ for (a) the VDSP of
$(V_a, V_b)=(10000, 50000)$,
and (b) the SDSP of $(S_a, S_b)=(10000,50000)$, 
results being successively shifted for clarity of figures ($a=0.01$ and $b=0.02$).
}
\label{fig12}
\end{figure}

\begin{figure}
\begin{center}
\includegraphics[keepaspectratio=true,width=100mm]{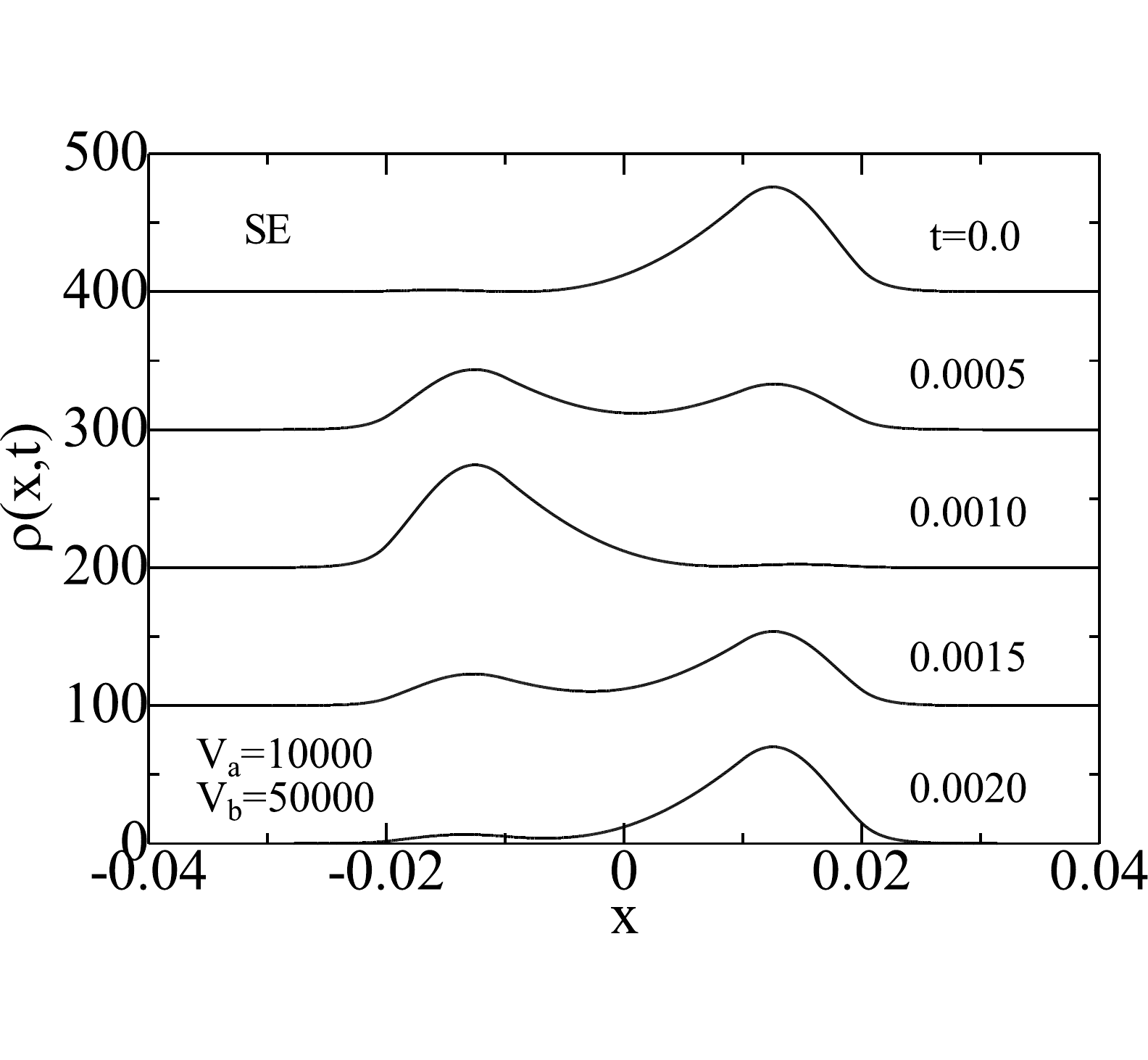}
\end{center}
\caption{
Time dependence of $\rho(x,t)$ of the Schr\"{o}dinger equation (SE)
for $(V_a,V_b)=(10000,50000)$, results being successively shifted for clarity of figure
($a=0.01$ and $b=0.02$).
}
\label{fig13}
\end{figure}

\subsection{Tunneling through the central barrier}
We may study the tunneling of a particle through the central potential barrier.
As an initial Gaussian-like wave packet, we assume a pair of the 
states for $n=1$ and $n=2$ as given by
\begin{eqnarray}
\Psi(x,t) &=& 
\left( {\begin{array}{*{20}c}
   \psi_{+}(x,t)  \\
   \psi_{-}(x,t)   \\
\end{array}} \right)
= \frac{1}{\sqrt{2}}
\left( {\begin{array}{*{20}c}
   \psi_{1+}(x)  \\
   \psi_{1-}(x)   \\
\end{array}} \right) \:e^{-i E_1 t/\hbar} +
\frac{1}{\sqrt{2}} 
\left( {\begin{array}{*{20}c}
   \psi_{2+}(x)  \\
   \psi_{2-}(x)   \\
\end{array}} \right) \:e^{-i E_2 t/\hbar},
\end{eqnarray}
where $\psi_{n \pm}(x)$ denotes spinor of the stationary wave function 
and $E_n$ signifies an eigenvalue of state $n$ ($=1,2$).
The time-dependent probability density $\rho(x,t)$ is given by
\begin{eqnarray}
\rho(x,t) &=& \vert \psi_+(x,t) \vert^2 + \vert \psi_-(x.t) \vert^2.
\end{eqnarray}
It is straightforward to calculate $\rho(x,t)$ because we have obtained
$\psi_{n \pm}(x)$ and $E_n$ for $n=1,2$ in the preceding Sec. III. 

Figure \ref{fig12}(a) shows $\rho(x, t)$ for the VDSP 
with $(V_a,V_b)=(10000,50000)$.
At $t=0.0$, $\rho(x, 0)$ consists of two Gaussian-like wave packets
because stationary wave functions of $\psi_{1\pm}(x)$ and $\psi_{2\pm}(x)$ 
have multiple nodes in Figs. \ref{fig5}(d) and \ref{fig5}(e).
The period of the oscillation is $T=2 \pi/\Delta E_{12}=0.000786$ 
for $\Delta E_{12}=E_2-E_1=7998$. 
The time dependence of $\rho(x,t)$ for the SDSP with 
$(S_a,S_b)=(10000,50000)$ shown in Fig. \ref{fig12}(b) has
the period of $T=0.0041$ for $\Delta E_{12}=1531$. 
The time dependence of $\rho(x,t)$ for the EDSP is similar to 
that for the SDSP (relevant result not shown).

Figure \ref{fig13} shows $\rho(x,t)$ of the Schr\"{o}dinger equation 
for the DSP with $(V_a, V_b)=(10000, 50000)$ which yields $T=0.00184$.
We note that $\rho(x,t)$ of the Schr\"{o}dinger equation
is similar to that for the SDSP in Fig. \ref{fig12}(b) 
but not to that for the VDSP in Fig. \ref{fig12}(a).  

\section{Conclusion}
Exact expressions of bound states for scalar and vector DSPs in the one-dimensional 
Dirac equation have been obtained with the use of 
the elegant transfer-matrix method.
Our calculations have shown that although results of the Dirac equation 
for scalar and vector DSPs reduce to those of the Schr\"{o}dinger 
in the nonrelativistic limit, they have the difference and similarity 
in general as follows:

\noindent
(i) The bound-state energy range of the Dirac equation for the VDSP 
is different from those for the SDSP and EDSP (Fig. \ref{fig3}),

\noindent
(ii) The bound-state energy $E_n$ has an approximate linear $n$ dependence
in the Dirac equation for adopted scalar and vector DSPs 
with small central potential barriers, while it is approximately given 
by $E_n \propto n^2$ in the Schr\"{o}dinger equation (Fig. \ref{fig10}), and

\noindent
(iii) The wave function and probability density of the Dirac equation 
for the VDSP are rather different from those of the Dirac equation 
for the SDSP and EDSP,
and also from those of Schr\"{o}dinger equation 
(Figs. \ref{fig5}, \ref{fig7}, \ref{fig9} and \ref{fig11}).

\noindent
As for the item (i), Eq. (\ref{eq:E1}) implies that the bound-state energy range
for the {\it vector} potential depends on the order of taking two limits of
$mc^2 \rightarrow \infty$ and $V_b \rightarrow \infty$.
If we first take the nonrelativistic limit of $mc^2 \rightarrow \infty$,
the bound-state range becomes $0 < E_s < V_b \rightarrow \infty$ ($E_s=E-mc^2$)
for infinite confining potential, 
in agreement with that of the Schr\"{o}dinger equation in Eq. (\ref{eq:E3}):
 $0 < E < V_b \rightarrow \infty$, as shown in Eq. (\ref{eq:C2}).
However, if we first take the limit of $V_b \gg mc^2$, 
the range for the bound state becomes $V_b-2mc^2 < E_s  < V_b$, which disagrees 
with the relevant result of the Schr\"{o}dinger equation. 
On the other hand, for the {\it scalar} potential, Eq. (\ref{eq:E2}) always
yields $0 < E_s < S_b$ for the positive eigenvalue in agreement 
with Eq. (\ref{eq:E3}) of the Schr\"{o}dinger equation.
This is consistent with Ref. \cite{Alberto96} 
in which Alberto, Fiolhals and Gil pointed out
that a calculation with the scalar potential avoids a difficulty 
realized with the vector potential, in studying a single square-well system with
an infinite confining  potential [Eq. (\ref{eq:B10d})].

Considering the fact that the double-well potential has been extensively
studied within the nonrelativistic treatment \cite{Thorwart01}, 
we expect that scalar and vector DSPs in the Dirac equation
play important roles in studying relativistic double-well systems, to which 
our method may be applied with various generalizations. 
Quite recently our nonrelativistic calculations have shown that an asymmetry 
in the double-well potential yields interesting quantum phenomena \cite{Hasegawa12,Hasegawa13}.
An application of the Dirac equation to an asymmetric
DSP is under consideration and results will be reported in a separate paper. 

\begin{acknowledgments}
This work is partly supported by
a Grant-in-Aid for Scientific Research from 
Ministry of Education, Culture, Sports, Science and Technology of Japan.  
\end{acknowledgments}

\appendix*

\section{Schr\"{o}dinger equation for the DSP}
\renewcommand{\theequation}{A\arabic{equation}}
\setcounter{equation}{0}

We obtain the bound-state solution of the Schr\"{o}dinger equation
\begin{eqnarray}
\left[- \frac{\hbar^2}{2m} \:\frac{d^2}{d x^2}  + V(x) \right] \psi(x) &=& E \psi(x),
\end{eqnarray}
for the DSP given by Eq. (\ref{eq:A1}).
Wave functions in five regions I$-$V are given by
\begin{eqnarray}
\psi_I(x) &=& A_1 \: e^{i q x}+ B_1 \: e^{- i q x} 
\hspace{1cm} \mbox{for $x \leq -b$}, \\
\psi_{II}(x) &=& A_2 \: e^{i k x}+ B_2\: e^{- i k x} 
\hspace{1cm} \mbox{for $-b < x \leq - a$}, \\
\psi_{III}(x) &=& A_3\: e^{i p x}+ B_3\: e^{- i p x} 
\hspace{1cm} \mbox{for $-a < x \leq a$}, \\
\psi_{IV}(x) &=& A_4\: e^{i k x}+ B_4\: e^{- i k x} 
\hspace{1cm} \mbox{for $a < x \leq b$}, \\
\psi_{V}(x) &=& A_5\: e^{i q x}+ B_5\: e^{- i q x} 
\hspace{1cm} \mbox{for $x > b$}, 
\end{eqnarray}
with
\begin{eqnarray}
k &=& \frac{\sqrt{2 m E}}{\hbar}, 
\label{eq:X1}\\
p &=& \frac{\sqrt{2m(E-V_a)}}{\hbar}, 
\label{eq:X2}\\
q &=& \frac{\sqrt{2m(E-V_b)}}{\hbar},
\label{eq:X3}
\end{eqnarray}
where $A_i$ ($B_i$) ($i=1-3$) denote magnitudes of wave functions traveling rightwards
(leftwards), and $m$ is mass of a particle. 
From the matching conditions for wave functions and their derivatives 
at the boundaries at $x= \pm a$ and $x= \pm b$, we obtain
\begin{eqnarray}
\left( {\begin{array}{*{20}c}
   {e^{ - iqb} } & {e^{iqb} }  \\
   {q\:e^{ - iqb} } & { - q\:e^{ iqb} }  \\
\end{array}} \right)\left( {\begin{array}{*{20}c}
   {A_1 }  \\
   {B_1 }  \\
\end{array}} \right) &=& \left( {\begin{array}{*{20}c}
   {e^{ - ikb} } & {e^{ikb} }  \\
   {k\:e^{ - ikb} } & { - k\:e^{ ikb} }  \\
\end{array}} \right)\left( {\begin{array}{*{20}c}
   {A_2 }  \\
   {B_2 }  \\
\end{array}} \right),
\label{eq:X4}
\end{eqnarray}

\begin{eqnarray}
\left( {\begin{array}{*{20}c}
   {e^{-ika} } & {e^{  ika} }  \\
   {k\:e^{-ika} } & { - k\: e^{  ika} }  \\
\end{array}} \right)\left( {\begin{array}{*{20}c}
   {A_2 }  \\
   {B_2 }  \\
\end{array}} \right) &=& \left( {\begin{array}{*{20}c}
   {e^{-ipa} } & {e^{  ipa} }  \\
   {p\: e^{-ipa} } & { - p\: e^{  ipa} }  \\
\end{array}} \right)\left( {\begin{array}{*{20}c}
   {A_3 }  \\
   {B_3 }  \\
\end{array}} \right),
\end{eqnarray}

\begin{eqnarray}
\left( {\begin{array}{*{20}c}
   {e^{ipa} } & {e^{ - ipa} }  \\
   {p\: e^{ipa} } & { - p\: e^{ - ipa} }  \\
\end{array}} \right)\left( {\begin{array}{*{20}c}
   {A_3 }  \\
   {B_3 }  \\
\end{array}} \right) &=& \left( {\begin{array}{*{20}c}
   {e^{ika} } & {e^{ - ika} }  \\
   {k\: e^{ika} } & { - k\: e^{ - ika} }  \\
\end{array}} \right)\left( {\begin{array}{*{20}c}
   {A_4 }  \\
   {B_4 }  \\
\end{array}} \right),
\end{eqnarray}

\begin{eqnarray}
\left( {\begin{array}{*{20}c}
   {e^{ikb} } & {e^{ - ikb} }  \\
   {k\: e^{ikb} } & { - k\: e^{ - ikb} }  \\
\end{array}} \right)\left( {\begin{array}{*{20}c}
   {A_4 }  \\
   {B_4 }  \\
\end{array}} \right) &=& \left( {\begin{array}{*{20}c}
   {e^{iqb} } & {e^{ - iqb} }  \\
   {q\: e^{iqb} } & { - q\: e^{ - iqb} }  \\
\end{array}} \right)\left( {\begin{array}{*{20}c}
   {A_5 }  \\
   {B_5 }  \\
\end{array}} \right).
\label{eq:X5}
\end{eqnarray}
Transfer matrix is given by
\begin{eqnarray}
\left( {\begin{array}{*{20}c}
   {A_1 }  \\
   {B_1 }  \\
\end{array}} \right) &=& \left( {\begin{array}{*{20}c}
   {T_{11} } & {T_{12} }  \\
   {T_{21} } & {T_{22} }  \\
\end{array}} \right)\left( {\begin{array}{*{20}c}
   {A_5 }  \\
   {B_5 }  \\
\end{array}} \right).
\end{eqnarray}

We note that Eqs. (\ref{eq:X4})-(\ref{eq:X5}) are equivalent 
to Eqs. (\ref{eq:A17})-(\ref{eq:A20}) for the Dirac equation
when we read $\alpha = k$, $\beta=q$ and $\gamma=p$.
The condition for the bound state is given by
\begin{eqnarray}
T_{11}&=&\frac{e^{2 i qb}}{16 k^2 q p}
\{
(k+p)^2  \left[ (k+q)^2 \:e^{2i[k(a-b)-pa]} 
- (k-q)^2 \:e^{-2i[k(a-b)- pa]} \right] \nonumber \\
&+& (k-p)^2 \left[ (k-q)^2 \:e^{-2i[k(a-b)+pa]} 
- (k+q)^2  \:\:e^{2i[k(a-b)+ pa]} \right] \nonumber \\
&+& 2 (k^2-q^2)(k^2-p^2) 
\left[\:e^{2i pa} - e^{-2ipa} \right]
\} = 0.
\label{eq:X6}
\end{eqnarray}
Bound states appear at
\begin{eqnarray}
0 < E < V_b,
\label{eq:X7}
\end{eqnarray}
for which $k$ is real and $q$ is purely imaginary:
plane waves in regions II and IV
and evanescent waves in regions I and V.

It is necessary to numerically solve the transcendental equation (\ref{eq:X6})
for given parameters of $m$, $V_a$, $V_b$, $a$, $b$.
Once an eigenvalue is obtained from Eq. (\ref{eq:X6}), matrix calculations
determine coefficients
of $A_i$, $B_i$ ($i=2$ to 4) and $B_1$ with $A_1=B_5=0$ for an assumed value of 
$A_5=C$ and $B_5=0$, as was made for the Dirac equation. 
The magnitude of $C$ is fixed by the normalization condition:
\begin{eqnarray}
\int_{-\infty}^{\infty} \vert \psi(x) \vert^2 \:dx=1. 
\end{eqnarray}

In the limit of $V_a=0$ or in the limit of $a=0$, Eq. (\ref{eq:X6}) becomes
\begin{eqnarray}
\frac{1}{4 k q} \left[ (q+k)^2 e^{-2 i k b}
- (q-k)^2 e ^{2i k b}\right] &=& 0,
\label{eq:X8}
\end{eqnarray}
which leads to the result for the single square-well potential
\begin{eqnarray}
\frac{2 k \kappa}{k^2-\kappa^2} &=& \tan(2 k b)
\hspace{1cm}\left( \kappa=\frac{\sqrt{2 m (V_b-E)}}{\hbar} \right).
\label{eq:X9}
\end{eqnarray}
In the limit of $V_b \rightarrow \infty$, 
Eq. (\ref{eq:X9}) yields the well-known result
\begin{eqnarray}
E_n &=& \frac{n ^2 \pi^2 \hbar^2}{8 m b^2} \hspace{1cm} \mbox{($n=1,2,\cdots$)}.
\label{eq:X10}
\end{eqnarray}


\end{document}